\documentclass[12pt]{article}
\usepackage{amsmath,amssymb}
\newcommand{\beq}{\begin{equation}}
\newcommand{\eeq}{\end{equation}}
\newcommand{\ba}{\begin{array}}
\newcommand{\ea}{\end{array}}
\newcommand{\bea}{\begin{eqnarray}}
\newcommand{\eea}{\end{eqnarray}}
\newcommand{\bean}{\begin{eqnarray*}}
\newcommand{\eean}{\end{eqnarray*}}

\newtheorem{theorem}{Theorem}[section]
\newtheorem{prop}[theorem]{Proposition}
\newtheorem{lem}[theorem]{Lemma}
\newtheorem{exe}[theorem]{Exercise}
\newtheorem{exa}[theorem]{Example}

\newtheorem{defi}[theorem]{Definition}
\newtheorem{remark}[theorem]{Remark}
\newenvironment{rem}{\begin{remark} \rm}{\end{remark}}

\newenvironment{exam}{\begin{exa} \rm}{\end{exa}}
\newtheorem{proof}{Proof.}
\newenvironment{pf}{\begin{proof} \rm}{\hfill$\square$ \end{proof}}
\newcommand{\CH}{{\cal H}}

\newcommand{\CZ}{{\cal Z}}

\newcommand{\CaC}{{\cal C}}

\newcommand{\CF}{{\cal F}}

\makeatletter
\@addtoreset{equation}{section}

\makeatother
\newcommand{\Ff}{{\mathsf F}}

\newcommand{\RR}{{\mathbb R}}
\newcommand{\CC}{{\mathbb C}}

\def\al{\alpha}

\def\la{\lambda} \def\La{\Lambda}
 
\def\om{\omega} 
\def\sig{\sigma}

\newcommand{\lmp}[3]{Lett. Math. Phys. {\bf #1} (#2), #3}

\newcommand{\jmp}[3]{Jour. Math. Phys. {\bf #1} (#2), #3}
\newcommand{\rref}[1]{(\ref{#1})} %puts parentheses around ref's

\def\dsl{\displaystyle}

\newcommand{\del}{{\partial}}

\def\parpo#1#2{\{#1,#2\}}
\def\parpol#1#2#3{\{#1,#2\}_\la^{#3}}
\def\mat2#1#2#3#4{{\left(\begin{array}{cc}#1 & #2\\ #3 & #4
\end{array}\right)}}

\def\mats2#1#2#3#4{{\left(\begin{array}{cc}#1 & #2\vspace{2truemm} \\ #3
&
#4
\end{array}\right)}}
\def\vec2#1#2{{\left[\begin{array}{c}
#1 \\ #2 \end{array}\right]}}

\def\ddd#1#2{\displaystyle{\frac{\partial #1}{\partial #2}}}

\newcommand{\Ha}[1]{H^{(#1)}}
\newcommand{\wHa}[1]{\widehat{H}^{(#1)}}
\def\wZHa#1#2#3{ \widehat{ Z_{#2}({H}^{(#1)}_{#3}) } }

\def\Nij{Nijenhuis}
\def\HJ{Hamilton--Jacobi\ }

\def\endpf{\par {}\hfill$QED$\\\medskip}

\def\ger{hierarch}
\def\var{manifold}

\def\bih{bi-Ham\-il\-tonian}
\def\varb{\bih\ \var}
\def\ham{Hamiltonian}
\def\vefi{vector field}

\def\parp{Poisson bracket}
\def\ger{hierarch}
\def\gerb{bi-Ham\-il\-ton\-ian hierarch}
\def\parpu{{\parpo{\cdot}{\cdot}}}
\def\tenp{Poisson tensor}
\def\tr{\mbox{tr}\,}

\def\parpopri#1#2{\{#1,#2\}'}

\def\syml{symplectic leaf}
\def\symls{symplectic leaves}
\def\syman{symplectic manifold}
\def\omnman{$\omega N$ manifold}

\def\dncoo{DN coordinates}
\def\wrt{with respect to}
\def\St{St\"ackel}
\def\stf{St\"ackel function}
\def\stsep{St\"ackel separable}
\def\stba{St\"ackel basis}

\def\bil{bi-Lagrangian}
\def\bilf{\bil\ foliation}

\newcommand{\stfg}{\stf\ generator}
\begin{document}
\baselineskip=18pt
%\begin{flushright}
%Ref. SISSA x/2002/FM
%\end{flushright}
\null\vskip 1.truecm
\begin{center}
{\huge \bf
Separation of variables for bi-Hamiltonian systems}
\end{center}
\vspace{0.8truecm}
\makeatletter
\begin{center}
{\large
Gregorio Falqui${}^a$
and Marco Pedroni${}^b$}
\\
\bigskip
${}^a$ SISSA, Via Beirut 2/4, I-34014 Trieste, Italy\\
E--mail: falqui@sissa.it\\
${}^b$ Dipartimento di Matematica, Universit\`a di Genova\\
Via Dodecaneso 35, I-16146 Genova, Italy\\
E--mail: pedroni@dima.unige.it\\
\vspace{1.truecm}
%{\large Version:\ \today}
\end{center}
\makeatother
\vspace{0.1truecm}
\begin{abstract}\noindent
We address
the problem of the separation of variables for the
Hamilton-Jacobi equation within the theoretical scheme of bi-Hamiltonian
geometry.
We use the properties of a special class of bi-Hamiltonian manifolds,
called
$\omega N$ manifolds, to give intrisic tests of separability (and
St\"ackel
separability)
for Hamiltonian systems. The separation variables are naturally
associated
with the geometrical structures of the $\omega N$ manifold itself. We
apply
these results to bi-Hamiltonian systems of the Gel'fand-Zakharevich type
and
we give explicit procedures to find the separated coordinates and the
separation relations.
\end{abstract}
%\newpage
\tableofcontents

\section{Introduction}

The technique of additive separation of variables for solving by
quadratures the Hamilton-Jacobi (HJ) equation is a very important tool
in analytical mechanics,
initiated by Jacobi and others
back in the nineteenth century
(see, e.g.,
\cite{Pars,DKN}). Following these classical works, an $n$-tuple
$(H_1,\ldots, H_n)$
of functionally independent Hamiltonians will be said to be separable in
a
set of canonical
coordinates $(q_1,\dots,q_n,p_1,\dots,p_n)$ if there exist $n$
relations,
called separation relations, of the form
\begin{equation}
\label{seprelint}
\phi_i(q_i,p_i,H_1,\dots,H_n)=0\ ,\quad i=1,\dots,n\ ,
\qquad\mbox{with }\det\left[\frac{\del\phi_i}{\del H_j}\right]
\not=0\> .
\end{equation}
The reason for this definition is that the stationary
Hamilton-Jacobi equations for the Hamiltonians $H_i$ can be
collectively solved by the additively separated complete integral
\begin{equation}\label{eq:i0}
W(q_1,\dots,q_n;\al_1,\dots,\al_n)=
\sum_{i=1}^n W_i(q_i;\al_1,\dots,\al_n)\>,
\end{equation}
where the $W_i$ are found by quadratures as the solutions of ordinary
differential equations.

One of the first systematic
results was found by Levi-Civita,
who provided, in 1904, a test for the separability of a given \ham\
in a given system of canonical coordinates. St\"ackel and Eisenhart
concentrated on \ham s quadratic in the momenta and orthogonal
separation variables. In particular, St\"ackel considered the \ham\
\[
H(q,p)=\frac12\sum g^{ii}(q){p_i}^2+V(q)
\]
and showed that
$H$ is separable in the coordinates $(q,p)$ if there exist an invertible
matrix $S(q)$ and a column vector $U(q)$
such that the $i$--th rows of $S$ and $U$ depend only on the
coordinate
$q_i$, and
$H$ is among the solutions $(H_1,\ldots, H_n)$ of the linear system
\[
\sum_{j=1}^n S_{ij}(q_i) H_j = \frac12 p_i^2-U_i(q_i)\ .
\]
These equations provide the separation relations for the
(commuting) \ham s $(H_1,\ldots, H_n)$.

With the works of Eisenhart, the theory of separation of variables was
inserted in the context of global Riemannian geometry, and this still
represents an active area of research, where the notions of Killing
tensor
and Killing web play a key role (see, e.g.,
\cite{Woodhouse,Kbook,Ben}).

Starting from the study of algebraic-geometric solutions of
(stationary reductions of) soliton equations and the
introduction of the concept of algebraic completely integrable system
\cite{AvM,DKN,NoVe}, separation of variables has received a renewed
attention
(see, e.g., \cite{FMcL,Ha1,GoNeRu,Hurtubise,Sk95}). This research
activity,
also connected with the theory of quantum
integrable systems,
deals
with Hamiltonian systems admitting a Lax representation
with spectral parameter and an $r$-matrix formulation. In this case,
the separation relations are provided by the spectral curve
\[
\det(\mu I-L(\la))=0
\]
associated with the Lax matrix $L(\la)$. Indeed, one can often find
canonical coordinates $(\la_1,\dots,\la_n,\mu_1,\dots,\mu_n)$ on the
phase space such that every pair $(\la_i,\mu_i)$ belongs to the
spectral curve. Since the \ham s are defined by the spectral curve,
they are separable in these coordinates.

The two classes of separable systems briefly recalled above
strongly suggest that a ``theory of separability''
should start from the following data,
\begin{enumerate}\label{data}
\item A class of symplectic manifolds $M$;
\item A class of canonical coordinates on $M$;
\item A class of Hamiltonian functions on $M$,
\end{enumerate}
and should provide
\begin{description}\label{fund.quest}
\item[{\rm a)}] separability test(s) to ascertain whether the HJ
equations
associated with the selected Hamiltonians
admit a complete integral which is additively separated in the
chosen coordinates;
\item[{\rm b)}] algorithms to compute the separation coordinates and
to exhibit the separation relations, so that the HJ equations can be
explicitly solved.
\end{description}
In the context of Riemannian geometry, the manifolds are cotangent
bundles of Riemannian manifolds, the coordinates are (fibered)
orthogonal coordinates, and the \ham s are quadratic in the momenta.
For Lax systems, roughly speaking, the manifolds are suitable
coadjoint orbits in loop algebras, the coordinates are the so-called
spectral Darboux coordinates \cite{Ha1}, possibly to be found using
the ``Sklyanin magic recipe'' \cite{Sk95}, and the separable \ham s are
the
spectral invariants.

The point of view herewith presented
is the following. The class of manifolds we will
consider
are particular \varb s, to be termed \omnman s, where one of the two
\parp s is nondegenerate and thus defines a symplectic form $\om$ and,
together with the other one, a {\em recursion operator\/}
$N$. The class of
coordinates, called {\em Darboux-Nijenhuis (DN) coordinates\/}, are
canonical \wrt\ $\om$ and diagonalize $N$.

The first result
is that an $n$-tuple $(H_1,\dots,H_n)$ of \ham s on $M$ (where
$n=\frac12 \dim M$) is separable in \dncoo\ if and only if they are in
involution \wrt\ both \parp s. This condition is clearly intrinsic,
i.e., it can be checked in any coordinate
system. A second result of the
present paper is that examples of separable systems on \omnman s are
provided by suitable reductions of \bih\ hierarchies, called
{\em Gel'fand-Zakharevich systems}. They are \bih\ systems defined
on a \varb\ $(M,\parpu,\parpu')$ by the coefficients of the Casimir
functions of the {\em Poisson pencil\/}
$\parpu_\la:=\parpu'-\la\parpu$. Such coefficients are in involution
\wrt\ both \parp s, and are supposed to be enough to define integrable
systems on
the \symls\ of $\parpu$. If there exists a foliation of $M$,
transversal to these \symls\ and compatible with the Poisson pencil
(in a suitable sense), then every \syml\ of $\parpu$ becomes an
\omnman, and the (restrictions of the)
GZ systems naturally fall
in the class of systems which are separable in DN
coordinates. For this reason, we can say that the Poisson
pencil separates its Casimirs.

The third result concerns the St\"ackel separability. With a slight
extension of the classical notion, we say that $(H_1,\ldots, H_n)$ are
St\"ackel separable if the
separation relations \rref{seprelint} are affine in the $H_i$:
\begin{equation}
\label{stseprelintro}
\sum_{j=1}^n S_{ij}(q_i,p_i) H_j-U_i(q_i,p_i)=0\ ,\qquad
i=1,\dots,n\ .
\end{equation}
In this case, the collection $(H_1,\ldots, H_n)$ is called a \stba. We
give
an intrinsic test for the St\"ackel
separability in \dncoo, which has a straightforward application to GZ
systems. This goes as follows. We notice that if $(H_1,\ldots, H_n)$ are
in
involution \wrt\ both \parp s (and therefore separable in \dncoo), then
there exists a matrix $F$ (depending on the choice of the $H_i$) such
that
\[
N^* dH_i=\sum_{j=1}^n F_{ij}dH_j\ .
\]
We prove that $(H_1,\ldots, H_n)$ is a \stba\ if and only if
\[
N^* dF_{ij}=\sum_{k=1}^n F_{ik}dF_{kj}\ .
\]

The geometric theory of separability we present in this paper
may be, in our opinion, regarded as an effective bridge between the
``classical'' and the ``modern'' aspects of the theory of separability.
More
evidence of this claim will be given in
\cite{fmp2}, where we will also show how to frame Eisenhart's theory
within
our approach, and discuss the problem of associating a Lax
representation to
GZ systems.

This paper is organized as follows. The first part (Section 2 to 5) is
devoted to
the geometry of separability on \omnman s. In
Section~\ref{sec:2} we will introduce the notion of \omnman\ and
we will study the \dncoo. Section~\ref{sec:4}
contains the main results about separability
on \omnman s, whereas in Section
\ref{sec:5} the \St\ separability is considered.
In Section~\ref{sec:6} we will come back
to DN coordinates, pointing out some
algorithms for their explicit computation.

In the second part of the paper we will turn our attention to GZ
systems. Section \ref{sec:7} deals with the particular case where
there is only one Casimir of the Poisson pencil (i.e., one \bih\
hierarchy), and contains the example of the 3-particle open Toda
lattice. This section is intended for an introduction to Section
\ref{sec:8},
where the general case is treated. We will give conditions
under which a \varb\ is foliated in \omnman s, and we will show that
the GZ systems are
separable in DN coordinates. Subsection \ref{subsec:8.3} is devoted to
the
\St\ separability of such systems. In Section \ref{sec:9} we will
show an efficient way to determine, in the \St\ separable case, the
separation relations for GZ systems.
Finally, we present an example in the loop algebra of
$\mathfrak{sl}(3)$.
\par\medskip\noindent
{\bf Acknowledgements.}
The results presented in this paper are a first
account of a long-standing
collaboration with Franco Magri, which we gratefully acknowledge. We
wish to
thank also Sergio Benenti, Boris Dubrovin, and John Harnad
for useful discussions. This work has been partially supported by
INdAM-GNFM
and
the Italian M.U.R.S.T. under the research project {\em Geometry of
Integrable Systems.}

\section{\bf{\omnman s}}\label{sec:2}
In this section we describe the manifolds where our (separable) systems
will
be defined. They are called {\em \omnman s\/}, since they are
Poisson-Nijenhuis (PN) manifolds \cite{KM,pondi,MMR} such that the first
Poisson structure is nondegenerate, and therefore defines a symplectic
form.
In turn, PN manifolds are particular instances of \varb s, i.e., smooth
(or
complex) manifolds $M$ endowed with a pair of of compatible \parp s,
$\parpu$ and $\parpu'$. This means that every linear combination of them
is
still a \parp.
\begin{defi}\label{def:1.1}
An \omnman\ is a \varb\ $(M,\parpu,\parpu')$ in which one of
the Poisson brackets (say, $\parpu$) is nondegenerate.
\end{defi}
Therefore, $M$ is endowed with a symplectic form $\om$ defined by
\begin{equation}
\label{omega}
\{f,g\}=\omega(X_f,X_g)\ ,
\end{equation}
where $X_f$ is the \ham\ \vefi\
associated with $f$ by means of $\parpu$. In terms of the \tenp\ $P$
corresponding to $\parpu$, viewed as a section of $\mbox{Hom}(T^*M,TM)$,
this simply means that $P$ is invertible and $\om$ is its inverse. Using
also the \tenp\ $P'$ associated with $\parpu'$, one can construct the
tensor
field $N:=P'P^{-1}$, of type $(1,1)$, to be termed {\em recursion
operator\/} of the \omnman\ $M$.
\begin{prop}\label{prop:2.1}
The \Nij\ torsion of $N$,
\begin{equation}\label{eq:2.2}
T(N)(X,Y):=[NX,NY]-N([NX,Y]+[X,NY]-N[X,Y])\ ,
\end{equation}
vanishes as a consequence of the compatibility between $P$ and $P'$.
\end{prop}
A proof of this well known fact can be found in \cite{pondi}.

There are two main sources of examples of \omnman.
The first one comes from classical mechanics. Let $Q$ be an
$n$-dimensional manifold endowed with a $(1,1)$ tensor field $L$ with
vanishing \Nij\ torsion, and let us
consider its cotangent bundle $T^*Q$ with the canonical \parp\ $\parpu$.
As
shown in \cite{IMM}, the vanishing of the \Nij\
torsion of $L$ entails that one can use it to define a second \parp\
$\parpu'$ on $T^*Q$ as
\[
\{q_i,q_j\}'=0\ ,\quad \{q_i,p_j\}'=-L_j^i\ ,\quad \{p_i,p_j\}'=
\left(\frac{\del L^k_j}{\del q_i}-\frac{\del L^k_i}{\del q_j}\right)p_k\
,
\]
where $(p_i,q_i)$ are fibered coordinates. This \parp\ is compatible
with
$\parpu$, so that the phase space $T^*Q$ becomes an \omnman, whose
recursion operator $N$ is the {\em complete lifting\/} \cite{Yano} of
$L$.

The second class of examples of \omnman s can be
obtained by reduction from a \varb\ $(M,P,P')$ where both
Poisson tensors are degenerate (see, e.g., \cite{Bedlewo}).

This happens, in particular, in the following situation.
Suppose that $P$ has constant corank
$k$,
that $\dim M=2n+k$, and that one can find a $k$-dimensional foliation
$\CZ$
of $M$ with the properties:
\begin{enumerate}
\item The foliation $\CZ$ is transversal to the symplectic foliation of
$P$;
\item The functions which are constant along $\CZ$ form a Poisson
subalgebra
of $(C^\infty(M),\parpu)$ and of $(C^\infty(M),\parpu')$, i.e., if $f$
and
$g$ are constant along $\CZ$, then the same is true for $\{f,g\}$ and
$\{f,g\}'$.
\end{enumerate}
Then any symplectic leaf $S$ of $\parpu$ inherits a \bih\ structure from
$M$. Moreover, the reduction of the first Poisson structure coincides
with
the symplectic form of $S$, so that $S$ is an \omnman. Such a procedure
is
one of the main topics of
the paper, and will be fully discussed in Section \ref{sec:8}, where we
will
also show that \bih\ systems on $M$ give rise to separable systems on
$S$.
The corresponding variables of separation are going to be introduced in the
next subsection.

\subsection{Darboux--\Nij\ coordinates}\label{sec:3}
In this subsection
we will describe a class of canonical coordinates on
\omnman s,
called Darboux--\Nij\ coordinates. They will play the important role of
variables of separation for (suitable) systems on \omnman s.
\begin{defi}
A set of local coordinates $(x_i, y_i)$ on an \omnman\ is called a set
of
{\em Darboux--\Nij\ (DN) coordinates\/} if they are canonical \wrt\ the
symplectic form $\om$,
\[
\omega=\sum_{i=1}^n dy_i\wedge dx_i\ ,
\]
and put the recursion operator $N$ in diagonal form,
\begin{equation}
\label{dn2}
N=\sum_{i=1}^n \la_i\left(
\ddd{}{x_i}\otimes d x_i+\ddd{}{y_i}\otimes d y_i\right)\ .
\end{equation}
This means that the only nonzero \parp s are
\[
\{x_i,y_j\}=\delta_{ij}\ ,\qquad \{x_i,y_j\}'=\la_i\delta_{ij}\ .
\]
\end{defi}
The assumption, contained in \rref{dn2}, that the eigenvalues $\la_i$ of
$N$
are (at least) double is not restrictive, since its eigenspaces have
even
dimension, equal to the dimension of the kernel of $P'-\la_i P$. For the
\omnman\ $T^*Q$ described in the previous section, it is easy to check
that
the eigenvalues of $L$ (if they are independent) and their conjugate
momenta
are DN coordinates. In order to ensure the existence of DN coordinates
on
more general \omnman s, we give the following
\begin{defi}\label{def:3.1}
A $2n$-dimensional \omnman\ $M$ is said to be {\em semisimple} if its
recursion operator $N$ has, at every point, $n$ distinct eigenvalues
$\la_1,\ldots, \la_n$.
It is called {\em regular} if the eigenvalues of $N$ are functionally
independent on $M$.
\end{defi}
It can be shown \cite{GZ93,Ma90,Turiel} that every point of a semisimple
\omnman\ has a neighborhood where \dncoo\ can be found, and that,
if the \omnman\ $M$ is regular, one half of these coordinates
are ``canonically'' provided by the recursion operator. Indeed, as a
consequence
of the vanishing of the \Nij\ torsion of $N$, the eigenvalues $\la_i$
always
satisfy
\[
N^* d\la_i=\la_i d\la_i\ ,
\]
where $N^*$ is the adjoint of $N$, and one has
\begin{prop}\label{prop:3.2}
In a neighborhood of a point of a regular \omnman\ where the eigenvalues
of
$N$ are distinct it is possible to find by quadratures $n$ functions
$\mu_1,
\ldots,
\mu_n$ that, along with the eigenvalues $\la_1,\ldots,\la_n$, are
\dncoo.
\end{prop}
Such coordinates will be called a set of
{\em special Darboux--\Nij\ (sDN) coordinates\/}. They will often be
used in
the sequel, because the $\la_i$ are simply the roots of the minimal
polynomial of $N$. Proposition \ref{prop:3.2} means also that every
regular
\omnman\ is locally equal to the ``lifted'' \omnman\ $T^*Q$ we have seen
in
Section \ref{sec:2}.

The distinguishing property of the pairs of \dncoo\ $(x_i,y_i)$,
and, a fortiori, of the ``special'' pairs $(\la_i,\mu_i)$, is that
their
differentials span an eigenspace of $N^*$, that is,
satisfy the equations
\begin{equation}\label{eq:3.1}
N^* dx_i=\la_i d x_i\ ,\qquad N^* dy_i=\la_i dy_i\ ,\qquad i=1, \ldots,
n\>.
\end{equation}
This leads us to the following
\begin{defi}
\label{def:stf}
A function $f$ on an \omnman\ is said to be a {\em \stf\/} (relative to
the
eigenvalue $\la_i$ of $N$) if
\begin{equation}
\label{stfun}
N^* df=\la_i df\ .
\end{equation}
\end{defi}
The following property of \stf s, which also explains their name, will
be
used many times in the rest of the paper.
\begin{prop}
\label{prop:exeig}
Let $M$ be a semisimple \omnman.
A function $f$ on $M$ is a \stf\ relative to $\la_i$ if and
only if, in any (some) system $(x_1,\dots,y_n)$ of DN coordinates, $f$
depends only on $x_i$ and $y_i$.
\end{prop}
{\bf Proof.} It is obvious that if $f=f(x_i,y_i)$ then $N^*df=\la_i
df$. Conversely, if \rref{stfun} holds, then $df$ belongs to the
$\la_i$-eigenspace of $N^*$, so that $df$ is a linear combination of
$dx_i$
and $dy_i$ and therefore $f$ depends only on $x_i$ and $y_i$.
\endpf

\section{Separability on \omnman s}\label{sec:4}
In Section 2
we have introduced a class of (symplectic) manifolds
and
we have selected a class of (canonical) coordinates on such manifolds.
Now we are going to characterize, from a geometric point of view,
those integrable \ham\ systems
on \omnman s which are separable in \dncoo. In the next section
we will consider the same problem for \St\ separability.

We recall that an $n$-tuple
$(H_1,\dots,H_n)$ of functionally independent Hamiltonians
on an \omnman\ $M$ is said to be separable in the DN coordinates
$(x_1,\dots,x_n,y_1,\dots,y_n)$ if there exist relations of the form
\begin{equation}
\label{seprel}
\phi_i(x_i,y_i,H_1,\dots,H_n)=0\ ,\quad i=1,\dots,n\ ,
\qquad\mbox{with }\det\left[\frac{\del\phi_i}{\del H_j}\right]
\not=0\> .
\end{equation}
It can be easily shown (e.g., via the Hamilton-Jacobi method) that this
entails the involutivity of the $H_i$.
Obviously enough, the separability property is not peculiar of the
specific
choice of the functions $H_i$. If $K_i=K_i(H_1,\dots,H_n)$ are
functions of the $H_i$, they are also separable according
to~\rref{seprel}.
So we see that the property~\rref{seprel} concerns the geometrical
features
of an integrable system, i.e., is to be regarded as a property of the
Lagrangian
distribution defined by the mutually commuting functions $H_i$. Thus one
can
say that the $H_i$ define a {\em separable foliation\/} of $M$.
According to
the following theorem, that will be proved during this section, the
separability property can be formulated in terms of the geometric
objects
$\om$ and $N$, or $\parpu$ and $\parpu'$, of the \omnman\ $M$.
\begin{theorem}\label{teo:4.1}
Let $M$ be a semisimple \omnman\ and let $(H_1,\ldots,H_n)$ be a set of
$n$
functionally independent Hamiltonians on $M$. Then the following
statements
are equivalent:

a) The foliation defined by $(H_1,\ldots,H_n)$ is separable in \dncoo\
(and
therefore Lagrangian \wrt\ $\om$);

b) The distribution tangent to the foliation defined by
$(H_1,\ldots, H_n)$ is Lagrangian \wrt\ $\omega$ and
invariant \wrt\ $N$;

c) The functions $(H_1,\ldots,H_n)$ are in bi-involution, i.e.,
$\{H_i,H_j\}=0$ and $\{H_i,H_j\}'=0$ for all $i,j$.
\end{theorem}
We will often refer to property c) by saying that the foliation defined
by
the $H_i$ is {\em \bil\/}. This is a fundamental property in our
approach to separability, and will be exploited especially in Sections
\ref{sec:7} and \ref{sec:8}. Incidentally, we notice that \bil\
foliations
play an important role
in the study of special K\"ahler manifolds \cite{Hitchin}.

Throughout the rest of the section $M$ will be a semisimple \omnman,
$(\la_1,\dots,\la_n)$ the eigenvalues of the recursion operator $N$, and
$(x_i,y_i)$ \dncoo\ on $M$. We begin with showing that the invariance
\wrt\
$N$ is a necessary condition for separability.

\begin{prop}
\label{prop:necforsep}
Let $(H_1,\dots,H_n)$ be functions on $M$ that are separable in DN
coordinates. Then the subspace spanned by $(dH_1,\dots,dH_n)$ is
invariant
\wrt\ $N^*$. More precisely, there exists a (simple) matrix $F$ with
eigenvalues $(\la_1,\dots,\la_n)$ such that
\begin{equation} \label{invcon}
N^* dH_i=\sum_{j=1}^n F_{ij}dH_j\ ,\qquad i=1,\dots,n\ .
\end{equation}
Consequently, the Lagrangian distribution defined by $(H_1,\dots,H_n)$,
which is spanned by the \ham\ \vefi s $X_{H_i}$, is invariant \wrt\ $N$.
\end{prop}
{\bf Proof.} Differentiate the relations \rref{seprel},
\begin{equation}
\frac{\del\phi_i}{\del x_i}dx_i+
\frac{\del\phi_i}{\del y_i}dy_i+
\sum_{j=1}^n \frac{\del\phi_i}{\del H_j}dH_j=0\ ,
\end{equation}
then apply $N^*$ to obtain
\begin{equation}
\frac{\del\phi_i}{\del x_i}\la_i dx_i+
\frac{\del\phi_i}{\del y_i}\la_i dy_i+
\sum_{j=1}^n \frac{\del\phi_i}{\del H_j}N^* dH_j=0\ .
\end{equation}
It follows that
\begin{equation}
\sum_{j=1}^n \frac{\del\phi_i}{\del H_j}N^* dH_j=-\la_i\left(
\frac{\del\phi_i}{\del x_i} dx_i+
\frac{\del\phi_i}{\del y_i} dy_i\right)=
\la_i\sum_{j=1}^n \frac{\del\phi_i}{\del H_j} dH_j\ ,
\end{equation}
that is, in matrix form,
\begin{equation}
J N^*dH=\La J dH\ ,
\end{equation}
where $J_{ij}=\frac{\del\phi_i}{\del H_j}$,
$dH=(dH_1,\dots,dH_n)^T$, $N^*dH=(N^*dH_1,\dots,N^*dH_n)^T$, and
$\La=\mbox{diag}(\la_1,\dots,\la_n)$. Therefore \rref{invcon} is
satisfied with $F=J^{-1}\La J$, and the eigenvalues of $F$ are
$(\la_1,\dots,\la_n)$. The final assertion easily follows.
\endpf
The matrix $F$ will be called the {\em control matrix\/}, \wrt\ the
basis
$(H_1,\dots,H_n)$, of the separable foliation.
\begin{prop}
\label{prop:inv-sep}
If $(H_1,\dots,H_n)$ define a distribution which is invariant \wrt\ $N$,
that is,
\begin{equation} \label{invcon2}
N^* dH_i=\sum_{j=1}^n F_{ij}dH_j\ ,\qquad i=1,\dots,n\ ,
\end{equation}
and the eigenvalues of $F$ are distinct, then the $H_i$ are separable in
\dncoo.
\end{prop}
{\bf Proof.} Since the eigenvalues of $F$ are distinct, they are the
eigenvalues
$(\la_1,\dots,\la_n)$ of $N$, so that there exists a matrix $S$ such
that
$F=S^{-1}\La S$. With $S$ we define the 1-forms
$\theta_i:=\sum_{j=1}^n S_{ij} dH_j$, for $i=1,\dots,n$. They are
eigenvectors of $N^*$, since
\begin{equation}
N^* \theta_i=\sum_{j=1}^n S_{ij} N^* dH_j=\sum_{j,k=1}^n S_{ij}
F_{jk} dH_k=\sum_{k=1}^n \la_i S_{ik} dH_k=\la_i\theta_i\ .
\end{equation}
Then there exist functions $L_i$ and $M_i$ such that
$\theta_i=L_i dx_i+M_i dy_i$, that is,
\begin{equation}
\sum_{j=1}^n S_{ij} dH_j-L_i dx_i-M_i dy_i=0\ .
\end{equation}
This means that $\dim\langle
dH_1,\dots,dH_n,dx_i,dy_i\rangle \le n+1$, so that
there exists a relation of the form \rref{seprel}, i.e., the functions
$(H_1,\dots,H_n)$ are separable in DN coordinates.
\endpf
In order to complete the proof of the equivalence between statements a)
and
b) of Theorem \ref{teo:4.1}, we need the following:
\begin{lem}
\label{lem:diseig}
If $(H_1,\dots,H_n)$ are independent functions in involution \wrt\ $\om$
such that \rref{invcon} holds, then the eigenvalues of $F$ are distinct.
\end{lem}
{\bf Proof.} Suppose that $N^* dH_i=\sum_{j=1}^n F_{ij}dH_j$, with
$\{H_i,H_j\}=0$ for all $i,j$. Since $F$ represents the
restriction of $N^*$ to $\langle dH_1,\dots,dH_n\rangle$,
%then the minimal polynomial of $F$ divides the one of $N^*$, so that
it is diagonalizable.
Thus, if $\la_i$ would be a double eigenvalue of $F$, the span
$\langle dH_1,\dots,dH_n\rangle$ would contain the
2-dimensional eigenspace spanned by $dx_i$ and $dy_i$.
But the involutivity of the $H_i$ would entail that $\{x_i,y_i\}=0$,
which is false.
\endpf
Relations \rref{invcon} may be called {\em generalized Lenard
relations\/} (and the functions $H_i$ fulfilling them a {\em Nijenhuis
chain\/}, as in ~\cite{FMT00}),
as enlightened by the following example.
\begin{exam}
\label{exam:tr}
If $H_k:=\frac1{2k}\mbox{tr}\,N^k=\sum_{j=1}^n \la_j^k$, then
$N^* dH_k=dH_{k+1}$ for $k=1,\dots,n-1$, that is, the Lenard relations
$P'dH_k=PdH_{k+1}$ hold. Moreover,
$dH_{n+1}=\sum_{j=1}^n c_j dH_{n+1-j}$,
where $\la^n-\sum_{j=0}^{n-1}c_{n-j}\la^j$ is the minimal polynomial
of $N$. Therefore,
condition \rref{invcon} is satisfied with
\begin{equation}\label{eq:compmat}
F=\left[\begin{array}{ccccc}
0 & 1 & 0 & \cdots & 0\\
0 & 0 & 1 & \cdots & 0\\
\vdots & \vdots & \vdots & \vdots & \vdots\\
0 & 0 & \cdots & \cdots & 1\\
c_n & c_{n-1} & \cdots & \cdots & c_1 \end{array}\right]\ .
\end{equation}
\end{exam}

\begin{rem}
It is well known that functions $H_i$ satisfying the Lenard relations
are in
involution \wrt\ both Poisson brackets, and so they provide a first
instance
of correspondence between invariant distributions and
bi-involutivity, which is at the same time trivial and paradigmatic.
\\
Indeed, it is trivial from the point of view of the theory of separation
of
variables, since such Hamiltonians are easily seen to depend only on
$(\la_1,\dots,\la_n)$ if the \omnman\ $M$ is regular and semisimple.
Then
the \HJ\ equations associated with the $H_i$ are trivially separable in
the
s\dncoo\ $(\la_1,\dots,\la_n,\mu_1,\dots,\mu_n)$. Nevertheless, it is
paradigmatic \wrt\ the issues of this paper. Indeed, the $H_i$ (that is,
the
$\la_i$) define a distinguished \bilf, called {\em principal
foliation\/},
which coincides with the canonical fibration $\pi:T^*Q\to Q$ of
classical
phase spaces when $T^*Q$ is the \omnman\ considered in Section
\ref{sec:2}.
However, there are in general \bilf s which are different from the
principal
one, as we will explicitly see in Section \ref{sec:7}. We are going to
show
that such foliations are characterized by the invariance \wrt\ $N$, so
that
they give rise to separable systems. This means that our theory deals
with
cases in which the Hamiltonians are not simply the
traces of the recursion operator. In other words, we will deal with
cases in
which the control matrix $F$ of equation \rref{invcon} need not be a
companion matrix of the form~\rref{eq:compmat}. Accordingly, the
separable
\vefi s we will consider are tangent to a \bilf, but they are not, in
general, \bih.
\end{rem}

\begin{prop}
\label{prop:inv-bilagr}
Let $(H_1,\dots,H_n)$ be independent functions on $M$. Then
\rref{invcon}
holds, with a matrix $F$ with distinct eigenvalues, if and only if the
functions $H_i$ are in bi-involution:
\begin{equation}
\label{bilagr}
\{H_i,H_j\}=\{H_i,H_j\}'=0\ ,\qquad
\mbox{for all $i,j=1,\dots,n.$}
\end{equation}
\end{prop}
{\bf Proof.} We know from Proposition \ref{prop:inv-sep} that condition
\rref{invcon}, with a simple matrix $F$, implies separability and
therefore
involutivity \wrt\ $\parpu$. Moreover,
\begin{equation}
\begin{array}{rl}
\{H_i,H_j\}' &=\langle dH_i,P' dH_j\rangle
=\langle dH_i,NP dH_j\rangle
=\langle N^* dH_i,P dH_j\rangle \\
&=\sum_{k=1}^n F_{ik} \{H_k,H_j\}\ .
\end{array}
\end{equation}
showing that $\{H_i,H_j\}'$ vanishes as well.

Conversely, suppose that $\{H_i,H_j\}=\{H_i,H_j\}'=0$ for all $i,j$.
Then
the foliation $\CH$ defined by the $H_i$ is Lagrangian \wrt\ $\parpu$,
and
\[
\langle N^* dH_i,P dH_j\rangle=\langle dH_i,NP dH_j\rangle
=\langle dH_i,P' dH_j\rangle=0\>.
\]
Thus, $N^* dH_i$ belongs, for every $i$, to
the annihilator of $\langle P\,dH_1,\dots,P\,dH_n\rangle$, which is
tangent to $\CH$, since $\CH$ is
Lagrangian. This shows that \rref{invcon} holds, and Lemma
\ref{lem:diseig}
entails that $F$ has distinct eigenvalues.
\endpf
Thus we have proved also the equivalence between b) and c) of Theorem
\ref{teo:4.1}.

\begin{rem}
One could also prove that a function $F$ is separable in DN coordinates
if and only if its Hamiltonian vector field $X_F$ is tangent to a
bi-Lagrangian foliation $\CH$.
The ``if'' part of this statement is a simple corollary of
Theorem~\ref{teo:4.1}. Indeed, let $\CH$ be defined by the functions
$(H_1,\ldots,H_n)$. Then $F$ is a function
of the $H_i$, since the distribution is Lagrangian, and one can find
other
$(n-1)$ functions $K_2,\ldots,K_n$ such that $\CH$ is defined by
$(F,K_2,\dots,K_n)$.
The ``only if'' part of this statement is deeper, and essentially gives
rise
to the intrinsic picture of the Levi-Civita conditions for separability,
to
be
fully discussed in \cite{fmp2}.
\end{rem}

Summing up, we have proved a criterion for the separability in DN
coordinates,
which can be tested {\em without\/} knowing explicitly these
coordinates.
Indeed, the statement \rref{bilagr} can be checked in any coordinate
system.
An important application of this criterion will be given in Section
\ref{sec:8}, where we will show that the \gerb ies on a \varb\ admitting
a
transversal distribution with the properties mentioned at the end of
Section \ref{sec:2} give rise to separable \ham\ \vefi s on the reduced
\omnman s.

\section{St\"ackel separability on \omnman s}
\label{sec:5}
The separability criteria of the previous section do not give explicit
information on the form of the separating relations \rref{seprel}. For
this
reason, in this section we will
concentrate on the more stringent notion of St\"ackel separability.
Recall
that $(H_1,\dots,H_n)$, independent functions on an \omnman, were
defined to
be
{\em \stsep\/} in the DN coordinates $(x_1,\dots,y_n)$ if there exist
relations of the form \rref{seprel},
given by affine equations in the $H_j$, that is,
\begin{equation}
\label{stseprel}
\sum_{j=1}^n S_{ij}(x_i,y_i) H_j-U_i(x_i,y_i)=0\ ,\qquad
i=1,\dots,n\ ,
\end{equation}
with $S$ an invertible matrix. In this case, we say that the $H_i$ are a
{\em \stba\/} of the (separable) foliation. The entries $S_{ij}$ and
$U_i$
depend only on $x_i$ and $y_i$, i.e., they are \stf s according to
Proposition \ref{prop:exeig}. Usually, $S$ is called a
{\em St\"ackel matrix\/}, and $U$ a {\em St\"ackel vector\/}. Notice
that
the definition of \St\ separability depends on the
choice of the $H_i$ defining the Lagrangian distribution. Indeed, if
$(H_1,\dots,H_n)$ are St\"ackel-separable, then
$K_i=K_i(H_1,\dots,H_n)$, for $i=1,\dots,n$, will not, in general,
fulfill relations of the form \rref{stseprel}. A natural problem, that
will
not be discussed in this paper, is to give a geometrical
characterization of
the Lagrangian foliations admitting a set of defining functions for
which
\St\ separability holds. Some results in this direction will be
presented in
\cite{fmp2}.

Now we will give a necessary and sufficient condition for
the St\"ackel separability in DN coordinates of a given $n$-tuple
$(H_1,\dots,H_n)$ of functions on an \omnman. We will also show that in
this
case one
can explicitly find the relations \rref{seprel} and has useful
information
to algebraically determine
the separation variables.

Suppose $(H_1,\dots,H_n)$ to be independent functions on a regular
semisimple
\omnman\ that are \stsep\ in the DN coordinates. Then we know from
Proposition \ref{prop:necforsep} that there exists a control matrix $F$,
with eigenvalues $(\la_1,\dots,\la_n)$, such that $N^*dH=FdH$.
Since Proposition \ref{prop:exeig} entails that $N^*dS=\La dS$ and
$N^*dU=\La dU$,
we can show:
\begin{prop}
\label{prop:necforstsep}
In the above-mentioned hypotheses, the matrix $F$ satisfies
\begin{equation}
\label{stsepcon}
N^* dF=FdF\ ,\qquad\mbox{that is, }\
N^*dF_{ij}=\sum_{k=1}^n F_{ik}dF_{kj}\quad\forall\
i,j=1,\dots,n\ .
\end{equation}
\end{prop}
{\bf Proof.} First we show that $F=S^{-1}\La S$. Indeed,
\begin{equation}
\begin{array}{rl}
SFdH &=SN^*dH=N^*\left[d(SH)-(dS)H\right]=N^*dU-(N^*dS)H\\
&=\La dU-\La dS\, H=\La SdH\ .
\end{array}
\end{equation}
Then we have
\begin{equation}
\begin{array}{rl}
N^*dF &=N^*d(S^{-1}\La S)=N^*(-S^{-1}dS\, S^{-1}\La S
+S^{-1}d\La\, S+S^{-1}\La dS)\\
&=-S^{-1}\La dS\, S^{-1}\La S
+S^{-1}\La d\La\, S+S^{-1}\La^2 dS\\
&=S^{-1}\La S(-S^{-1}dS\, S^{-1}\La S
+S^{-1}d\La\, S+S^{-1}\La dS)=FdF\ ,
\end{array}
\end{equation}
and the proof is complete.
\endpf
Condition \rref{stsepcon} is also sufficient for the
St\"ackel separability, as shown in the following:
\begin{theorem}
\label{thm:stsep}
Let $(H_1,\dots,H_n)$ be independent functions, defining a \bilf\ on a
regular semisimple \omnman. If the control matrix $F$ fulfills
\rref{stsepcon}, then:
\begin{enumerate}
\item The left eigenvectors of
$F$, if suitably normalized, form a St\"ackel matrix. More precisely,
if $S$ is a matrix such that $F=S^{-1}\La S$, and such that
in every row of $S$ there is an entry equal to 1, then $S$ is a
St\"ackel matrix in DN coordinates $(x_1,\dots,y_n)$;
\item The functions $(H_1,\dots,H_n)$ are \stsep\ in DN coordinates.
\end{enumerate}
\end{theorem}
{\bf Proof.} From \rref{stsepcon} we have that
\begin{equation}
\begin{array}{rl}
& N^*(-S^{-1}dS\, S^{-1}\La S + S^{-1}d\La\, S+S^{-1}\La dS)\\
& \qquad = S^{-1}\La S(-S^{-1}dS\, S^{-1}\La S + S^{-1}d\La\,
S+S^{-1}\La
dS)\ ,
\end{array}
\end{equation}
that is,
\begin{equation}
N^*(-dS\, S^{-1}\La S + \La dS)=
\La (-dS\, S^{-1}\La S + \La dS)\ ,
\end{equation}
or $(-N^* dS + \La dS)F=\La (-N^* dS + \La dS)$. Hence the
$j$-th row of $(-N^* dS + \La dS)$ is a left eigenvector of $F$,
relative to $\la_j$. This entails that it is proportional to the
$j$-th row of $S$, i.e., there exists a 1-form $\al_j$ such that
\begin{equation}
\label{propej}
e_j (-N^* dS + \La dS)=\al_j e_j S\ ,
\end{equation}
where $e_j$ is the $j$-th row vector of the standard basis.
Multiplying equation \rref{propej} by $e_k^T$, where $S_{jk}=e_j S
e_k^T=1$, we obtain $\al_j=0$, so that
\begin{equation}
N^* dS=\La dS\ .
\end{equation}
In components, this reads $N^* dS_{jk}=\la_j dS_{jk}$, which
implies (see Proposition \ref{prop:exeig}) that $S_{jk}$ depends only
on $x_j$ and $y_j$, i.e., $S$ is a St\"ackel matrix. Finally, the fact
that
$U:=SH$ is a St\"ackel vector follows from
\begin{equation}
N^* dU=N^* (dS\, H+S dH)=\La dS\, H+SF dH=\La (dS\, H+S dH)=\La dU \ .
\end{equation}
This completes the proof.
\endpf
The results obtained so far can be summarized in the following
statements. An $n$-tuple of functions $(H_1,\dots,H_n)$ in involution is
separable (in DN coordinates)
if and only if the span of their differentials is invariant for $N^*$.
Let $F$ be the matrix which represents (the restriction of) $N^*$ on
such a span. Then equation \rref{stsepcon} represents a test for the
{\em St\"ackel\/} separability %in DN coordinates
of the $H_i$. Once this test is passed, the St\"ackel matrix is easily
constructed as a (suitably normalized) matrix that diagonalize $F$,
and the separation procedure can be quite explicitly performed.
Therefore,
in
our setting the Hamiltonians provide their St\"ackel matrix as well as
the separation relations \rref{stseprel}.

We end this section with the following comment on the intrinsic meaning
of
the \St\ separability conditions~\rref{stsepcon}. It is known
\cite{FrNi}
that, as a consequence of
the vanishing of the \Nij\ torsion of $N$, the de Rham complex
of $M$ is
endowed with a second derivation $d_N$, which is defined to be the
unique
(anti)derivation \wrt\ the wedge product extending
\begin{equation}\label{diffn}
\begin{array}{rcl}
d_N f (X) &=& df(NX)=N^*df(X)\\
d_N \theta (X,Y) &=& X(\theta(NY))-Y(\theta(NX))-\theta([X,Y]_N)\ ,
\end{array}
\end{equation}
where $f$ is a function, $\theta$ is a 1-form, $X$, $Y$ are vector
fields on $M$, and
\[
[X,Y]_N=[NX,Y]+[X,NY]-N[X,Y]\ .
\]
This differential is a cohomology operator ($d_N^2=0$) and anticommutes
with
the usual exterior derivative $d$. One notices that the invariance
condition~\rref{invcon} can be equally be written,
in matrix notation, as
\begin{equation}
\label{eq:4.x}
d_N H=F\,dH\>.
\end{equation}
Imposing the condition $d_N^2=0$ on this equation, taking into
account the anticommutativity of $d$ and $d_N$, and translating back
$d_Nf=N^*df$ if $f$ is a function on $M$, one gets
\begin{equation}
\label{eq:4.y}
\left(N^*dF-F dF\right)\wedge dH=0\> .
\end{equation}
So we see that the \St\ separability conditions \rref{stsepcon}
are nothing but a
``strong'' solution of the equations imposed on the control matrix $F$
by
the cohomological condition $d_N^2=0$.

\section{Special DN coordinates}\label{sec:6}
In this section we will discuss the problem of explicitly finding sets
of
special DN coordinates on an \omnman\ $M$. We
assume that $M$ be regular and complex, so that the eigenvalues
$(\la_1,\dots,\la_n)$ of $N$ can be used as (half of the) coordinates on
$M$. We know that in a neighborhood of a point where the $\la_i$ are
distinct there exist functions $(\mu_1,\dots,\mu_n)$ forming with the
eigenvalues a system of \dncoo, and that the $\mu_i$ can be computed by
quadratures.
However, they can often be found in an algebraic way, as we will see
below.
We divide our argument in three main points.

We start remarking that there are simple conditions to be checked the
$\mu_i$ in order to ensure that they form with the
$\la_i$ a set of DN coordinates. To this aim, we observe that the
$\mu_i$ must fulfill two kinds of requirements:
\begin{enumerate}
\item They have to be \stf s, that
is, they must satisfy $N^* d\mu_i=\la_i d\mu_i$;
\item They have to fulfill the canonical commutation relations
\wrt\ the first \parp: $\{\la_i,\mu_j\}=\delta_{ij}$,
$\{\mu_i,\mu_j\}=0$.
\end{enumerate}
In principle, these conditions require the computation of the
$\la_i$. We will show that this can be avoided, and that a
smaller number of equations must be checked.
The first step is to notice that, once conditions 1 are
satisfied, conditions 2 can be replaced with the $n$ equations
\begin{equation}
\label{mucon}
\{\la_1+\cdots+\la_n,\mu_i\}=1\ ,
\end{equation}
which do not require the computation of the $\la_i$, but only of
their sum, that is, $c_1:=\frac12 \mbox{tr}\, N$ and, consequently,
of the \ham\ vector field $Y:=-P dc_1=\sum_{i=1}^n \frac{\del}{\del
\mu_i}$.
Indeed, suppose that $\mu_j$ be a \stf, and observe that
\begin{equation}
\begin{array}{rl}
\la_i\{\la_i,\mu_j\}&=
\la_i\langle d\la_i,Pd\mu_j\rangle=
\langle \la_i d\la_i,Pd\mu_j\rangle=
\langle N^* d\la_i,Pd\mu_j\rangle\\
&=\langle d\la_i,N Pd\mu_j\rangle=
\langle d\la_i,PN^* d\mu_j\rangle=
\langle d\la_i,\la_j P d\mu_j\rangle\\
&=\la_j\{\la_i,\mu_j\}\ ,
\end{array}
\end{equation}
so that $\{\la_i,\mu_j\}=0$ if $i\not=j$. Then equation
\rref{mucon} becomes $\{\la_i,\mu_i\}=1$. In the same way
one shows that $\{\mu_i,\mu_j\}=0$. Hence, in order to find the
$\mu_i$ coordinate we have to look
for a \stf\ (relative to $\la_i$) such that \rref{mucon}
holds.

The second point starts from the following idea, which will be
extensively
used in the part of the paper dealing with Gel'fand--Zakharevich
systems.
Let us consider the minimal polynomial
\begin{equation}
\label{eq:delta}
\Delta(\la)=\la^n-(c_1\la^{n-1}+c_2\la^{n-2}+\cdots c_n)
\end{equation}
of $N$. Using the Newton formulas relating the traces of the powers of
$N$
and
the coefficients $c_i$ of $\Delta(\la)$, one easily verifies that the
latter
satisfy
\begin{equation}
\label{eq:frobechain}
\begin{split}
& N^* dc_i=dc_{i+1}+c_i dc_1\>,\qquad i=1,\ldots, n-1\\
& N^* dc_{n-1}=d c_n\>.\end{split}
\end{equation}
These relations are equivalent to the following equation for the
polynomial $\Delta(\la)$,
\begin{equation}
\label{eq:fropol}
N^* d \Delta(\la)=\la d \Delta(\la)+\Delta(\la)d c_1\>.
\end{equation}
Relations of this kind are very interesting for our purposes. For
instance,
it holds:
\begin{prop}\label{prop:eeg}
Let $f(x;\la)$ be a function defined on $M$, depending on an additional
parameter $\la$.
Suppose that there exists a 1-form $\alpha_f$ such that
\begin{equation}
\label{eq:exeiggene}
N^* d(f(x;\la))=\la d(f(x;\la)) +\Delta(\la)\alpha_f\>.
\end{equation}
Then, the function $f_i$ defined by $f_i(x):=f(x;\la_i(x))$, i.e., the
evaluation of $f(x;\la)$ on $\la=\la_i$, is a \stf\ relative to $\la_i$.
\end{prop}
{\bf Proof.}
The differential of $f_i$ equals
\begin{equation}
\label{eq:g5.x1}
d f_i(x)=df(x;\la){\big\vert_{\la=\la_i}}+{\frac{\del f(x;\la)}
{\del \la}}\big\vert_{\la=\la_i} d\la_i\ ,
\end{equation}
where, in the term
$d \left(f(x;\la)\right){\big\vert_{\la=\la_j}}$, one treats $\la$ as a
parameter. Applying the adjoint of the recursion operator we get
\begin{equation}
\label{eq:g5.x2}
N^*d \left({f}(x;\la_i)\right)=N^*d \left(f(x;\la)\right)
{\big\vert_{\la=\la_i}}+\la_i\dsl{\frac{\del {f}(x;\la)}
{\del \la}}\big\vert_{\la=\la_i} d\la_i\ ,
\end{equation}
whence the assertion, since $\Delta(\la_i)=0$.
\endpf
\begin{defi} We will call a function on $M$ depending on the additional
parameter $\la$ a {\em \stf\ generator} if it satisfies
\rref{eq:exeiggene} with a suitable 1-form $\alpha_f$.
\end{defi}
\begin{lem}
The space of \stfg s is closed under sum and multiplication, and is
invariant \wrt\ the action of $Y$. If $f$ is a \stfg\ and $g$ is a
function
of one variable, then $g\circ f$ is a \stfg.
\end{lem}
{\bf Proof}. The only assertion whose proof is not straightforward is
the
invariance \wrt\ $Y$. This follows from the fact, already noticed in
Example
\ref{exam:tr}, that $Y$ is a \bih\ \vefi; hence $L_Y(N^*)
=L_Y(P^{-1}P')=0$ and, consequently, $L_Y(\la_i)=0$.
\endpf
It is clear that if $f_i$ is a \stf\ relative to $\la_i$ for
$i=1,\dots,n$,
then there exists a \stfg\ $f(x;\la)$ such that $f_i=f(x;\la_i)$, e.g.,
the
interpolating polynomial. In terms of the generator, condition
\rref{mucon}
can be written as
\begin{equation}
\label{eq:nmu}
Y(f(x;\la))=1\qquad \mbox{for $\la=\la_i$, $i=1,\dots,n$.}
\end{equation}
For further use, we state and prove the following:
\begin{prop}
\label{prop:perlemu}
Let $f$ be a \stfg, and suppose that for $n\ge 1$ the action of $Y$ on
$f$
closes, that is, the relation
\begin{equation}
Y^n(f)=\sum_{j=0}^{n-1} a_j Y^j(f)\ ,
\end{equation}
with $Y(a_j)=0$, holds.
Then equation~\rref{eq:nmu} can be algebraically solved.
\end{prop}
{\bf Proof}.
There are two cases: in a first instance, suppose that,
actually, $Y$ is nilpotent, that is, $Y^n(f)=0$ is
satisfied for some $n\ge 1$ (whilst $Y^{n-1}(f)\neq 0$).
Then it is easily seen that $Y^{n-2}(f)/Y^{n-1}(f)$ is a \stfg\
fulfilling
\rref{eq:nmu}. \\
On the contrary, if $(a_0,\dots,a_{n-1})\not=(0,\ldots, 0)$, then the
matrix
$A$ representing the action of $Y$ on
$\Phi:=(f,Y(f),\dots,Y^{n-1}(f))^T$
has at least one nonzero
eigenvalue $\nu$, which is a solution of $\nu^n=\sum_{i=0}^{n-1}
a_j\nu^j$.
Let $w=(w_0,\dots,w_{n-1})$ be a (left) eigenvector of $A$ relative to
$\nu$, e.g., the one given by $w_{n-1}=1$ and
$w_k=\nu^{n-k-1}-\sum_{j=0}^{n-k-2} a_{k+j+1}\nu^j$ for $k=0,\dots,n-2$.
Then
\begin{equation}
\label{eq:munorm}
Y\left(\frac1{\nu}\log\sum_{j=0}^{n-1} w_j Y^j(f)\right)=1\ .
\end{equation}
Indeed, $\sum_{j=0}^{n-1} w_j Y^j(f)=w\Phi$ and
\[
Y\left(w\Phi\right)=wA\Phi=\nu w\Phi\ ,
\]
implying \rref{eq:munorm}.
\endpf

These arguments reveal a further important aspect of \St\ separability
within our approach to separation of variables. Indeed, the
condition of \St\ separability, whose intrinsic form is
given by equation~\rref{stsepcon}, entails that the matrix of the
(suitable
normalized) eigenvectors
of the control matrix $F$ is a \St\ matrix, that is, its columns are
\stf s of $N^*$. Since we have shown that a way to algebraically find
the
$\mu_i$ coordinates is to find \stf s (or \stfg s) and to combine them
in
order to fulfill equation~\rref{eq:nmu}, we see
that, in the \St\ case, the \ham s themselves may algebraically provide
the
coordinates in which the corresponding flows can be separated.

\section{Separability on odd-dimensional \varb s}
\label{sec:7}
This section starts the second (and more applicative) part of the paper,
in
which we will use the results of Sections
\ref{sec:4} and \ref{sec:5} to discuss the separability of a specific
family
of integrable systems. They are defined on a class
of \varb s, known in the literature
as complete torsionless \varb s of pure Kronecker type
(see \cite{GZ99,Pa99} and the references quoted therein).

In this section we will consider the simplest case, corresponding to
generic
odd-dimensional \varb s (while in Section \ref{sec:4} we studied the
case of
regular \omnman s, which are generic
even-dimensional \varb s). Their Poisson tensors have maximal rank. The
more
general case will be treated (with detailed proofs) in the next section.

Let $(M,P,P')$ be a $(2n+1)$-dimensional \varb, and let the rank of $P$
be
equal to $2n$. Suppose that the {\em Poisson pencil\/} $P_\la:=P'-\la P$
has
a polynomial Casimir function
\[
H(\la)=\sum_{i=0}^n H_i\la^{n-i}\ .
\]
This amounts to saying that the functions $(H_0,\dots,H_n)$, which we
assume
to be functionally independent, form a \gerb y, starting from a Casimir
$H_0$ of $P$ and terminating with a Casimir of $P'$,
\begin{equation}
\label{lenrel}
P\,dH_0=0\ ,\quad P\,dH_{i+1}=P'\,dH_i\ ,\quad P'\,dH_n=0\ .
\end{equation}
In particular, they are in involution \wrt\ $\parpu$ and $\parpu'$. If
$dH_0\ne 0$ at every point of $M$, then the symplectic foliation of $P$
is
simply given by the level surfaces of $H_0$. The restrictions of
$(H_1,\dots,H_n)$ to a \syml\ $S$ of $P$ form an integrable system (in
the
Arnold-Liouville sense). The corresponding \ham\ \vefi s are the
restrictions to $S$ of $X_i:=P\,dH_i$, where $i=1,\dots,n$.

At this point it is natural to wonder whether the \bih\ structure of $M$
can
give information on the separability of the (restrictions of the) \ham s
$(H_1,\dots,H_n)$. More concretely, one can try to induce an $\omega N$
structure on $S$ in order to apply the separability theorems of Sections
\ref{sec:4} and \ref{sec:5}. As anticipated in Section \ref{sec:2}, this
can
be done if there exists a vector field $Z$ which is transversal to the
symplectic foliation of $P$ and fulfills the following condition:

C) if $F$, $G$ are functions on $M$ which are invariant for $Z$, that
is,
$Z(F)=Z(G)=0$, then $\{F,G\}$ and $\{F,G\}'$ are also invariant.
\par\noindent
In this case, any \syml\ of $P$ inherits a \bih\ structure from $M$.
Clearly, the first reduced bracket is the one associated with the
symplectic
form of $S$, so that $S$ is an \omnman.

In the following section we will prove that, if $Z$ is normalized in
such a
way that $Z(H_0)=1$, condition C) takes the infinitesimal form
\begin{equation}
\label{infform}
L_Z P=0\ ,\qquad L_Z P'=Y\wedge Z\ ,
\end{equation}
for a suitable vector field $Y$. In this case there is a useful form of
the
reduced \parp s $\parpu_S$ and $\parpu'_S$ on the \syml\ $S$. If $f$ and
$g$
are functions on $S$, by definition $\{f,g\}_S$ and $\{f,g\}'_S$ should
be
computed by taking local extensions of $f$ and $g$ which are invariant
along $Z$. But one can avoid the use of invariant functions and consider
arbitrary extensions $F$, $G$. Then
\[
\begin{array}{rcl}
\{f,g\}_S &=& \{F,G\}\\
\{f,g\}'_S &=& \{F,G\}'+X'(F)Z(G)-X'(G)Z(F)\ ,
\end{array}
\]
where $X':=P'\,dH_0=P\,dH_1$ and the right-hand sides of the previous
equations are implicitly understood to be restricted to $S$. These
equations
show that the restrictions of $(H_1,\dots,H_n)$ to $S$ are in
bi-involution,
and then separable in \dncoo\ because of Theorem \ref{teo:4.1}. We are
going
to show that they are even \stsep, by computing their control matrix $F$
and
checking that it satisfies the condition $N^*dF=F\,dF$.

To this purpose, we notice that the Lenard relations \rref{lenrel} on
$M$
give rise to the equations
\begin{eqnarray}
\label{lenrelsus1}
N^* d{\hat H}_i &=& d{\hat H}_{i+1}-\widehat{Z(H_i)} d{\hat H}_1\
,\qquad
i=1,\dots,n-1\ , \\
\label{lenrelsus2}
N^* d{\hat H}_n &=& \phantom{d{\hat H}_{i+1}}-\widehat{Z(H_n)} d{\hat
H}_1\
,
\end{eqnarray}
where $N$ is the recursion operator of the \omnman\ $S$ and
$\hat{\ }$ denotes the restriction to $S$. Therefore, the control
matrix of $({\hat H}_1,\dots,{\hat H}_n)$ is given by a
single Frobenius block:
\begin{equation}
\label{eq:g2.2}
F=\left[\begin{array}{cccccc}
-Z(H_1)&1 &0&&\dots&0\\
-Z(H_2)&0&1&&&\\
\vdots&&&\ddots&&\\
\vdots&&&&&1\\
-Z(H_{n})&&&&&0\end{array}
\right]\ .
\end{equation}
So we see that the (restriction to the symplectic leaf $S$) of the
functions $c_i=-Z(H_i)$
are the coefficients of the characteristic polynomial of the matrix $F$,
that is, the coefficients of the minimal polynomial
of the recursion operator $N$,
$\Delta(\la)=\la^n-(c_1\la^{n-1}+\cdots+c_n)$.
Recalling that the coefficients of the minimal polynomial of $N$ satisfy
\begin{equation}\label{eq:rec.c}
N^* d c_i=d c_{i+1}+c_{i} d c_1\ ,\quad N^*d c_n=c_{n} d c_n\ ,
\end{equation}
we see that the condition $N^*dF=F\,dF$ for the St\"ackel
separability of the \ham s is automatically verified. Hence we have
proven
\begin{theorem}
The \ham s of a corank-1 torsionless GZ system are
St\"ackel separable in DN coordinates.
\end{theorem}
It is worthwhile to notice that the examples previously considered in
the literature within the theory of quasi-bi-Hamiltonian systems (see,
e.g.,
\cite{Bl98,mt97,ybzeng}) fall into this class. The link with the
classical
St\"ackel-Eisenhart theory of separation of variables is discussed in
\cite{IMM}.

We remark that the \vefi\ $Y$ appearing in \rref{infform} can be chosen
to
be tangent to $S$. In this case, $Y=P\,d(Z(H_1))=-P\,dc_1$, so that its
restriction to $S$ is the \vefi\ we used in the previous section to
determine the $\mu_i$ coordinates. (This explains why we made use of the
same notation).

Now we will write the separation equations for the GZ \ham s. The
\St\ matrix $S$, being the (normalized) matrix of the left eigenvectors
of
$F$, is easily seen to be the Vandermonde-like matrix
\[
S=\left[\begin{array}{cccc}
\la_1^{n-1} & \cdots & \la_1 & 1\\
\vdots & \cdots & \vdots & \vdots\\
\la_n^{n-1} & \cdot & \la_n & 1
\end{array}\right]\ ,
\]
where the $\la_i$ are the eigenvalues of $N$, i.e., the roots of
$\Delta(\la)$.
Therefore, the separation relations take the form
\begin{equation}
\label{seprelco1}
{\hat H}_1\la_i^{n-1} +{\hat H}_2\la_i^{n-2} \cdots +
{\hat H}_n=U_i(\la_i,\mu_i)\ ,
\end{equation}
where $(\la_1,\dots,\la_n,\mu_1,\dots,\mu_n)$ are special \dncoo\ on $S$
and
the $U_i$ are the entries of the \St\ vector. Such entries can be
explicitly
computed once we have the map sending the \dncoo\ to the
corresponding point of $S$, as we will check in the example of the
3-particle nonperiodic Toda lattice.

Another way to arrive at the separation equations is to multiply
\rref{lenrelsus1} by $\la^{n-i}$ and then to add to
\rref{lenrelsus2}. The result is
\[
N^*\,d{\hat H}(\la)=\la\,d {\hat H}(\la)-\Delta(\la)d {\hat H}_1\ ,
\]
meaning that ${\hat H}(\la):=\sum_{i=1}^n {\hat H}_i\la^{n-i}$ is a
\stfg\
according to Proposition \ref{prop:eeg}. Thus, in \dncoo,
${\hat H}(\la_i)=
{U_i}(\la_i,\mu_i)$, which coincides with \rref{seprelco1}. We stress
that
${\hat H}(\la)$, being a \stfg, can be in some cases used to determine
the
$\mu_i$ coordinates. Instances of this situation are provided by the
Toda
lattice, as discussed in \cite{creta}, and by the stationary reductions
of
the KdV hierarchy \cite{FMPZ2}. Here we will present the example of the
3-particle nonperiodic Toda lattice.

\begin{exam}
The Hamiltonian of the system is
\begin{equation}
\label{hamTod}
H_{\mbox{\scriptsize Toda}}=\frac12\sum_{i=1}^3
{p_i}^2+\sum_{i=1}^2\exp(q^i-q^{i+1})\ .
\end{equation}
As usual (see, e.g.,~\cite{Fla}, and \cite{FMcL} for the
separability),
one introduces the
``Flaschka-Manakov coordinates'' $(a_1,a_2,b_1,b_2,b_3)$, where
\[
b_i=p_i\ ,\qquad a_i=-\exp(q^i-q^{i+1})\ ,
\]
and consider the manifold
$M=(\CC^*)^2\times \CC^3$, or $M={\RR_{>0}}^2\times \RR^3$.
We endow it with the Poisson pencil $P_\la=P'-\la P$ given by (see,
e.g.,
\cite{MoPi} and references cited therein)
\begin{equation}
P_\la=
\left [\begin {array}{ccccc} 0&-a_{{1}}a_{{2}}&\left (b_{{1}}-\lambda
\right )a_{{1}}&\left (\lambda-b_{{2}}\right )a_{{1}}&0
\\\noalign{\medskip}a_{{1}}a_{{2}}&0&0&\left (b_{{2}}-\lambda\right )a
_{{2}}&\left (\lambda-b_{{3}}\right )a_{{2}}\\\noalign{\medskip}\left
(\lambda-b_{{1}}\right )a_{{1}}&0&0&a_{{1}}&0\\\noalign{\medskip}
\left (b_{{2}}-\lambda\right )a_{{1}}&\left (\lambda-b_{{2}}\right )a_
{{2}}&-a_{{1}}&0&a_{{2}}\\\noalign{\medskip}0&\left (b_{{3}}-\lambda
\right )a_{{2}}&0&-a_{{2}}&0\end {array}\right ]\ .
\end{equation}
It has a polynomial Casimir $H(\la)=H_0\la^2+H_1\la+H_2$, where
\begin{eqnarray*}
H_0 &=& b_{{1}}+b_{{2}}+b_{{3}} \\
H_1 &=&
-(b_{{1}}b_{{2}}+b_{{2}}b_{{3}}+b_{{3}}b_{{1}}+a_{{1}}+a_{{2}})\\
H_2 &=& b_{{1}}b_{{2}}b_{{3}}+a_{{1}}b_{{3}}+a_{{2}}b_{{1}}\ .
\end{eqnarray*}
The \ham\ \rref{hamTod} is related to the coefficients of $H(\la)$ by
$H_{\mbox{\scriptsize Toda}}=H_1+\frac12 H_0$.
There are two nontrivial flows, given by:
\begin{eqnarray*}
X_1 &=& P_0\,dH_1=a_1(b_1-b_2)\frac{\del}{\del a_1}+
a_2(b_2-b_3)\frac{\del}{\del a_2}+a_1\frac{\del}{\del b_1}+
(a_2-a_1)\frac{\del}{\del b_2}-a_2\frac{\del}{\del b_3}\\
X_2 &=& P_0\,dH_2=a_1[a_2+b_3(b_2-b_1)]\frac{\del}{\del a_1}+
a_2[a_1+b_1(b_3-b_2)]\frac{\del}{\del a_2}-
a_1b_3\frac{\del}{\del b_1}\\
&&+
(a_1b_3-a_2b_1)\frac{\del}{\del b_2}+a_2b_1\frac{\del}{\del b_3}\ .
\end{eqnarray*}
The symplectic leaves of $P$ are the level surfaces of $H_0$, so that
they
can be parametrized by $(a_1,a_2,b_1,b_2)$. A possible choice for the
normalized
transversal vector field is $Z=\frac{\del}{\del b_3}$, because
$Z(H_0)=1$
and
\[
L_Z P=0\ ,\qquad L_Z P'=Y\wedge Z\ ,
\]
with $Y=a_2\frac{\del}{\del a_2}$. Since $Y(H_0)=0$, we know that $Y=P
d(Z(H_1))=-P\,dc_1$. If $S$ is a \syml\ of $P$, the reduced \bih\
structure
on $S$ is simply obtained by removing the last row and the last column
of
$P_\la$:
\[
P_S=\left[\begin{array}{cccc}
0 & 0 & a_1 & -a_1 \\
0 & 0 & 0 & a_2 \\
-a_1 & 0 & 0 & 0 \\
a_1 & -a_2 & 0 & 0
\end{array}\right]\ ,\quad
P'_S=\left[\begin{array}{cccc}
0 & -a_1 a_2 & a_1 b_1 & -a_1 b_2 \\
a_1 a_2 & 0 & 0 & a_2 b_2 \\
-a_1 b_1 & 0 & 0 & a_1 \\
a_1 b_2 & -a_2 b_2 & -a_1 & 0
\end{array}\right]\ .
\]
For completeness, we display
recursion operator,
\[
{N}={P}'_S P_S^{-1}=\left[\begin{array}{cccc}
b_1 & a_1(b_1-b_2)/a_2 & a_1 & a_1 \\
0 & b_2 & -a_2 & 0 \\
0 & a_1/a_2 & b_1 & 0 \\
-1 & -a_1/a_2 & 0 & b_2
\end{array}\right]\ ,
\]
whose minimal polynomial is
\[
\Delta(\la)=\la^2 +Z(H_1)\la +Z(H_2)=
\la^2 -(b_1+b_2)\la +a_1+b_1b_2\ .
\]
The coordinates $\la_1$, $\la_2$ are its roots.

The restrictions of $H_1$ and $H_2$ to the \syml\ $H_0=c$ are
\begin{eqnarray*}
{\hat H}_1 &=& -c(b_{{1}}+b_{{2}})+{b_{{1}}}^2+{b_{{2}}}^2+b_{{1}}
b_{{2}}-a_{{1}}-a_{{2}}\\
{\hat H}_2 &=& c(a_1+b_1b_2)-(a_1+b_1b_2)(b_{{1}}+b_{{2}})+
a_{{2}}b_{{1}}\ .
\end{eqnarray*}
We know that ${\hat H}(\la):={\hat H}_1 \la+{\hat H}_2$ is a \stfg, and
that
the separation equations are ${\hat H}(\la_i)=U(\la_i,\mu_i)$, for
$i=1,2$.
To write them explicitly, we need the form of the $\mu_i$. They can be
found
using Proposition \ref{prop:perlemu} and the fact that
\[
Y^2 ({\hat H}(\la))=Y({\hat H}(\la))\ .
\]
This entails that $f(\la):=\log Y(\hat H(\la))$ satisfies $Y(f(\la))=1$,
so
that, according to the results of Section \ref{sec:6},
\[
\mu_i=\log Y(\hat H(\la_i))=\log(a_2 b_1-a_2\la_i)\ ,\qquad i=1,2\ ,
\]
form with the eigenvalues $\la_1$ and $\la_2$ of $N$ a set of (special)
\dncoo.

Finally, using the expression of $(a_1,a_2,b_1,b_2)$ in terms of the
\dncoo\
one can easily find the separation relations
\[
\hat H(\la_i)={\la_i}^3+\exp \mu_i-c{\la_i}^2\ ,\qquad i=1,2\ ,
\]
leading to the solution by quadratures of the Hamilton-Jacobi equations
for
${\hat H}_1$ and ${\hat H}_2$.

We notice that the ``change of variables''
$(a_i, b_i)\mapsto (\la_i,\mu_i)$
is not the lift of a point
trasformation on the configuration space; thus, there is no
contradiction
with the results of \cite{BMcLS}, stating that it is impossible to
separate
the 3-particle Toda lattice with point tranformations.
\end{exam}

\section{Separability of Gel'fand--Zakharevich systems}
\label{sec:8}

In this section we will generalize (and give proofs of) the results of
the
previous section to the case of corank $k$. As we will see, the picture
outlined in the previous section still holds good. The only relevant
difference concerns the \St\ separability, which is no longer valid in
general, but requires an additional assumption on the \ham s.

We consider a \varb\ $(M,P,P')$ admitting $k$ polynomial Casimir
functions
of the Poisson pencil $P_\la=P'-\la P$,
\begin{equation}
\label{polcas}
\Ha{a}(\la)=\sum_{i=0}^{n_a} \Ha{a}_{i}\lambda^{n_a-i}\ ,\qquad
a=1,\dots,k\
,
\end{equation}
such that $n_1+n_2+\cdots+n_k=n$, with $\dim M=2n+k$, and such that the
differentials of the coefficients $\Ha{a}_i$ are linearly independent on
$M$. The $\Ha{a}_i$, for a fixed $a$, form a \gerb y and, in particular,
$\Ha{a}_0$ (resp.\ $\Ha{a}_{n_a}$) is a Casimir of $P$ (resp.\ $P'$). We
assume that the corank of $P$ is exactly $k$, so that the $\Ha{a}_0$,
for
$a=1,\dots,k$, are a maximal set of independent Casimirs of $P$.
The collection of the $n$ \bih\ vector fields
\begin{equation}
\label{eq:01}
X^{(a)}_i=P\, d\Ha{a}_{i}=P'\,d\Ha{a}_{i-1}\ ,\quad
i=1,\dots,n_a,\quad
k=1,\dots,a\ ,
\end{equation}
associated with the Lenard sequences defined by the Casimirs
$\Ha{a}$ is called the {\em Gel'fand--Zakharevich (GZ) system\/},
or the {\em axis\/}, of the \varb\ $M$.
Since standard arguments from the theory of Lenard--Magri chains show
that
all the coefficients $\Ha{a}_{i}$ pairwise commute with respect to both
$\parpu$ and $\parpopri{\cdot}{\cdot}$, we have
\begin{prop}\label{prop:AL-inte}
Let $S$ be a symplectic leaf of $P$, that is, a
$2n$--dimensional submanifold defined by
$\Ha{1}_0=c_1,\ldots,\Ha{k}_0=c_k$. Then the vector fields
$X^{(a)}_i$
of the Lenard sequences associated with the polynomial Casimirs
\rref{polcas} of $\parpol{\cdot}{\cdot}{}$
on $M$ define a completely integrable Hamiltonian system
on $S$.
\end{prop}
We call the family $\{\wHa{a}_i\mid i=1,\dots,n_a, k=1,\dots,a\}$
of the restrictions to $S$ of the coefficients of the $\Ha{a}$ the {\em
GZ
basis\/} of the \syml\ $S$. The lagrangian foliation defined by the GZ
basis
will be referred to as the {\em GZ foliation\/}
of $S$.

In the following subsection we will give sufficient conditions so that a
\syml\ $S$ of $P$ inherits an $\omega N$ structure from the \bih\
structure
of $M$. Then we will come back to the integrable system described in
the
previous proposition and we will discuss its separability in \dncoo.

\subsection{The induced $\omega N$ structure}
Our strategy to induce on a \syml\ $S$ of $P$ a second \parp\ which is
compatible with the ``canonical'' one is based on the geometrical
considerations already mentioned at the end of Section 2.
We suppose that there exists a $k$-dimensional foliation
$\CZ$ of $M$ such that
\begin{description}
\item [{\rm C1)}] the foliation $\CZ$ is transversal to the symplectic
foliation of $P$;
\item [{\rm C2)}] the functions that are constant on $\CZ$ form a
Poisson
subalgebra \wrt\ both $\parpu$ and $\parpu'$.
\end{description}
Thus $S$ has a (projected) \bih\ structure. The projection of $\parpu$
coincides with the symplectic structure $\parpu_S$ of $S$, while the
projection of $\parpu'$ defines a second \parp\ $\parpu'_S$ on $S$.
Since
the compatibility between $\parpu_S$ and $\parpu'_S$ is guaranteed by
the
fact that the whole pencil $\parpu_\la$ is projectable on $S$, we have
endowed $S$ with an $\omega N$ structure. We will suppose it to be a
regular
\omnman, in order to apply (in the open dense set where the eigenvalues
of
$N$ are distinct) the results of Section \ref{sec:4} and \ref{sec:5},
leaving the discussion of the
problem of finding the conditions on
$(M,P,P')$ and $\CZ$ ensuring the regularity of $S$ for future work.

Let $(Z_1,\dots,Z_k)$ be local \vefi s spanning the distribution
tangent
to $\CZ$. Because of the transversality condition, we can always
normalize
these \vefi s \wrt\ the Casimirs
$\Ha{a}_0$ of P:
\begin{equation}
\label{eq:nrmZ}
Z_b(\Ha{a}_0)=\delta^a_{b}\> .
\end{equation}
In terms of these generators, the projectability condition takes a very
concise form, as shown in
\begin{prop}
\label{prop:zvefi}
~\linebreak\noindent
1. The normalized \vefi s $Z_a$ locally generating $\CZ$ are symmetries
of
$P$,
\begin{equation}
\label{zap}
L_{Z_a}(P)=0\ ,
\end{equation}
and satisfy
\begin{equation}
\label{zapp}
{L}_{Z_a} P'=\sum_b Y^b_a\wedge Z_b\ ,
\end{equation}
where $Y^b_a=P\,d(Z_a(\Ha{b}_1))=[Z_a,P'\,d\Ha{b}_0]=
[Z_a,X_1^{(b)}]$.
\par\noindent
2. Viceversa, suppose that there exists a $k$-dimensional integrable
distribution on $M$ which is transversal to the symplectic leaves of $P$
and
such that \rref{zap} and \rref{zapp} hold for a suitable local basis
$(Z_1,\dots,Z_k)$ of the distribution
(and for suitable \vefi s $Y^b_a$). Then the integral foliation of the
distribution satisfies the projectability requirements C1) and C2), so
that
every \syml\ of $P$ becomes an \omnman. Moreover, if the $Z_a$ are
normalized, then they commute.
\end{prop}
{\bf Proof}.
First of all, we recall (\cite{vaismanbook}, p.\ 54) that the condition
that
the functions constant along $\CZ$
form a Poisson subalgebra with respect to $\parpu$
is equivalent to the assertion that the following equations hold,
\begin{equation}
\label{eq:a1}
{L}_{Z_a} P=\sum_{b=1}^k W^b_a\wedge Z_b\ ,
\end{equation}
for some vector fields $W^b_a$. This entails the validity of assertion
2,
except the commutativity of the \vefi s $Z_a$, normalized according to
\rref{eq:nrmZ}, that can be proved as follows. The integrability of the
distribution implies that there are functions $\phi^c_{ab}$ such that
\[
\left[Z_a,Z_b\right]=\sum_{c=1}^k \phi^c_{ab}Z_c\ ,
\]
and evaluating this relation on the Casimirs $\Ha{d}_0$ of $P$ we easily
see
that $\phi^c_{ab}=0$.

In order to prove assertion 1, we notice that the vector fields
${W}^b_a$ are not uniquely defined, and can be taken to be tangent to
the
\symls\ of $P$. This is accomplished by changing
\[
W^b_a\mapsto W^b_a-\sum_{c=1}^k W^b_a(\Ha{c}_0)Z_c\ .
\]
Indeed,
\begin{eqnarray*}
\sum_{b=1}^k (W^b_a-\sum_{c=1}^k W^b_a(\Ha{c}_0)Z_c)\wedge Z_b &=&
\sum_{b=1}^k W^b_a\wedge Z_b-\sum_{b,c=1}^k W^b_a(\Ha{c}_0)Z_c\wedge
Z_b\\
&=&\sum_{b=1}^k W^b_a\wedge Z_b
\end{eqnarray*}
since $L_{Z_a}\langle d\Ha{c}_0,P\,d\Ha{d}_0\rangle=0$ and
\rref{eq:a1} implies that
\[
W^d_b(\Ha{c}_0)=W^c_b(\Ha{d}_0)\ .
\]
Thus the \vefi s $W^b_a$ in \rref{eq:a1} can be chosen in such a way
that
$W^b_a(\Ha{c}_0)=0$. Now, deriving the relation $P d\Ha{c}_0=0$
along $Z_a$ one obtains that the normalized vector fields $W^b_a$
vanish, so
that,
indeed, the vector fields $Z_a$ are symmetries of $P$.

As far as the second Poisson tensor $P'$ is concerned, in the same way
we
can show that there exist vector fields $Y^b_a$ tangent to the \symls\
of
$P$ such that
\begin{equation}
\label{eq:a2}
{L}_{Z_a} P'=\sum_{b=1}^k Y^b_a\wedge Z_b\ .
\end{equation}
By deriving the relation $P' d\Ha{c}_0=X_1^{(c)}$
with respect to $Z_a$, one has that
\[
Y^c_a=[Z_a,X_1^{(c)}]=L_{Z_a}(P\,d\Ha{c}_1)=P d( Z_a(\Ha{c}_1))\ .
\]
This completes the proof.
\endpf

In the sequel we will always suppose that the normalization conditions
\rref{eq:nrmZ} on the transversal \vefi s $Z_a$ and the tangency
conditions
on the $Y^b_a$ are satisfied. For the sake of simplicity, we will also
assume that the $Z_a$ are defined on the whole manifold $M$, or at least
in a
tubular neighborhood of $S$. Next we give a useful formula for the
(second)
reduced \parp\ on $S$.
\begin{prop}
\label{prop:redparp}
Let $f$, $g$ be functions on a \syml\ $S$ of $P$, and $F$, $G$ arbitrary
extensions of $f$, $g$ to $M$. Then
\begin{eqnarray}
\label{firred}
\{f,g\}_S &=& \{F,G\}\\
\label{secred}
\{f,g\}'_S &=&
\{F,G\}'+\sum_{a=1}^k\left(X_1^{(a)}(F)Z_a(G)-X_1^{(a)}(G)Z_a(F)\right)\
,
\end{eqnarray}
where $X_1^{(a)}=P'\,d\Ha{a}_0=\{\Ha{a}_0,\cdot\}'$.
\end{prop}
{\bf Proof.} The symplectic leaf $S$ is given by the equations
$\Ha{a}_0=c^a$, for $a=1,\dots,k$, where the $c^a$ are suitable
constants.
The first formula simply says that $\{\cdot,\cdot\}_S$ corresponds to
the
symplectic structure of $S$. The second formula follows from the remark
that
$\tilde F:=F-\sum_{a=1}^k Z_a(F)\left(\Ha{a}_0-c^a\right)$ coincides
with
$F$ and fulfills $Z_b(\tilde F)=0$ on $S$. Hence it can be used to
compute
$\{f,g\}'_S$, giving \rref{secred}.
\endpf

\begin{rem}
The projectability conditions we have imposed in order to endow a fixed
\syml\ $S$ with an $\om N$ structure can be weakened in the following
way.
We can consider a distribution transversal to $TS$ and defined only at
the
points of $S$, generated by a family of \vefi s $(Z_1,\dots,Z_k)$,
normalized as $Z_a(\Ha{b}_0)=\langle d\Ha{b}_0,Z_a\rangle=\delta_a^b$.
Then
we introduce, according to \rref{secred}, a composition law $\parpu'_S$
on
$C^\infty(S)$ and we look for conditions ensuring that it is a \parp,
compatible with $\parpu_S$. One can show
\cite{Bedlewo} that $\parpu'_S$ is a \parp\ if and only if
\begin{equation}\label{eq:sc2}
\sum_{a=1}^k X^{(a)}_1\wedge
\left(L_{Z_a}(P')+\sum_{b=1}^k[Z_a,X'_b]\wedge
Z_b\right)+\frac12 \sum_{a,b=1}^k X^{(a)}_1\wedge X^{(b)}_1\wedge
[Z_a,Z_b]=0
\end{equation}
at the points of $S$. In this case, the two Poisson brackets are
compatible
if and only if
\begin{equation}\label{3.17}
\sum_{a=1}^k X^{(a)}_1\wedge L_{Z_a}(P)=0
\end{equation}
at the points of $S$. Hence, the requirements \rref{zap} and
\rref{zapp}, on the whole manifold $M$, are very ``strong'' solutions
for
\rref{eq:sc2} and \rref{3.17}. Finally, we mention that the reduction
process presented in this remark does not fit in the Marsden-Ratiu
scheme
\cite{MR86}, whereas the one based on C1) and C2) clearly does.
\end{rem}

\subsection{Separability and the control matrix}
\label{subsec:8.2}

After endowing any \syml\ $S$ of $P$ with an $\om N$ structure, we can
reconsider the GZ foliation of $S$ and prove its separability in \dncoo.
Notice that (see also below) the restrictions to $S$ of the \bih\ \vefi
s
$X_i^{(a)}$ are {\em not\/} \bih\ \wrt\ the $\om N$ structure of $S$.
This
is due to the fact that this structure is obtained by means of a {\em
projection}, while the \ham\ are {\em restricted\/} to $S$.

We suppose that $(Z_1,\dots,Z_k)$ are \vefi s on $M$, fulfilling the
hypotheses of part 2 of Proposition \ref{prop:zvefi} and normalized,
i.e.,
$Z_a(\Ha{b}_0)=\delta_a^b$. Then the expressions \rref{firred} and
\rref{secred} of the reduced \parp s immediately show that the
restrictions
of $\Ha{a}_i$ to $S$ are in
bi-involution. Therefore, they are separable in \dncoo.

\begin{theorem}
\label{teo:gzsep}
The GZ foliation of $S$ is separable in DN coordinates.
\end{theorem}

Using once more Theorem \ref{teo:4.1}, we can conclude that the
distribution
tangent to the GZ foliation is invariant \wrt\ the recursion operator
$N$.
We are going to describe the form of the associated control matrix,
which
will be needed to discuss the \St\ separability of the GZ basis.

Let $g$ be any function on $S$ and let $G$ be an extension of $g$ to
$M$.
Using \rref{secred} and the Lenard relations on the $\Ha{a}_i$, we have
\begin{equation}\label{eq:g1.5}
\begin{aligned}
\{\wHa{a}_i,g\}'_S &= \{\Ha{a}_i,G\}'+\sum_{b=1}^k
\left(X^{(b)}_1(\Ha{a}_i) Z_b(G)-
X^{(b)}_1(G) Z_b(\Ha{a}_i)\right)\\
&= \{\Ha{a}_{i+1},G\}-\sum_{b=1}^k Z_b(\Ha{a}_i) \{H^{(b)}_1,G\}\ ,
\end{aligned}
\end{equation}
where we have put $\Ha{a}_{n_a+1}:=0$. Therefore, for all $g\in
C^\infty(S)$,
\begin{equation}\label{eq:g1.5b}
\{\wHa{a}_i,g\}'_S = \{\wHa{a}_{i+1},g\}_S-\sum_{b=1}^k
\widehat{Z_b(\Ha{a}_i)} \{\wHa{b}_1,g\}_S\ ,
\end{equation}
or, in terms of the (reduced) \tenp s $P_S$ and $P'_S$,
\begin{equation}\label{eq:g1.5c}
P'_S\,d\wHa{a}_i=P_S\,d\wHa{a}_{i+1}-\sum_{b=1}^k
\widehat{Z_b(\Ha{a}_i)}
P_S\,d\wHa{b}_1\ .
\end{equation}
Hence, we can conclude that
\begin{equation}\label{FforGZ}
N^*\,d\wHa{a}_i=d\wHa{a}_{i+1}-\sum_{b=1}^k \widehat{Z_b(\Ha{a}_i)}
d\wHa{b}_1\
\end{equation}
and read the form of the control matrix
$F$ associated with the GZ basis. Indeed, if we order the $n$ functions
of
the GZ basis as
\begin{equation}\label{eq:g1.seq}
\wHa{1}_1,\wHa{1}_2,\ldots,\wHa{1}_{n_1},\wHa{2}_1,\ldots,
\wHa{k}_{n_k}\ ,
\end{equation}
then we realize that $F$ has a $k\times k$ block form,
\begin{equation}
\label{eq:g1.10}
F=\left[
\begin{array}{cccc}
\CF_{1}& \CaC_{1,2} & \cdots &\CaC_{1,k}\\
\CaC_{2,1}&\CF_{2}&\cdots & \CaC_{2,k}\\
\vdots &&&\vdots\\
\CaC_{k,1}&&&\CF_{k}
\end{array}\right]\ ,
\end{equation}
with $\CF_{a}$ an $n_a\times n_a$ square matrix of Frobenius type of
the
form
\begin{equation}\label{eq:fromat2}
\CF_{a}=\left[\begin{array}{cccccc}
-\widehat{Z_a(\Ha{a}_1)}&1 &0&&\dots&0\\
-\widehat{Z_a(\Ha{a}_2)}&0&1&&&\\
\vdots&&&\ddots&&\\
\vdots&&&&&1\\
-\widehat{Z_a(\Ha{a}_{n_a})}&&&&&0\end{array}
\right]
\end{equation}
and $\CaC_{a,b}$ a rectangular matrix with $n_a$ rows and $n_b$ columns
where only the first column is nonzero:
\begin{equation}\label{eq:fromat}
\CaC_{a,b}=\left[\begin{array}{cccc}
-\widehat{Z_b(\Ha{a}_1)}&0 &\dots&0\\
-\widehat{Z_b(\Ha{a}_2)}&0 &\dots&0\\
\vdots&\vdots&&\vdots\\
-\widehat{Z_b(\Ha{a}_{n_a})}&0&\dots&0\end{array}
\right]\ .
\end{equation}
\begin{rem}
\label{rem:yvf}
The \vefi\ $Y$, defined in Section \ref{sec:6} as the \ham\ \vefi\
associated with $-\frac12 \tr N$ by the first Poisson structure, can be
obtained in the present setting by restricting to $S$ the \vefi\
$\sum_{a=1}^k Y_a^a$. Indeed,
\[
\sum_{a=1}^k Y_a^a=P\,d\left(\sum_{a=1}^k Z_a(\Ha{a}_1)\right)\ ,
\]
and using \rref{eq:g1.10} we have
$\sum_{a=1}^k Z_a(\Ha{a}_1)=-\tr F=-\frac12 \tr N$.
\end{rem}

Thus, we have seen that GZ systems on \varb s admitting a suitable
transversal foliation provide examples of non trivial (but still
somewhat
special) Hamiltonian
systems for which the separability condition in DN coordinates holds,
that
is,
they provide interesting examples of control matrices,
discussed in Section~\ref{sec:4}.
Such matrices were introduced (in the specific example of a stationary
reduction of the Boussinesq equation)
in~\cite{FMT00}.

\subsection{St\"ackel separability of GZ systems}
\label{subsec:8.3}

Let us now consider the St\"ackel (i.e., linear) separability of
GZ systems. We have seen that the invariance with
respect to $N$ of the Lagrangian distribution defined by the restricted
Hamiltonians $\wHa{a}_i$ is a consequence of the Lenard recursion
relations
on $M$, and
that the nontrivial coefficients in $F$ are given
by the deformations of the polynomial Casimirs along the normalized
generators $Z_a$ of the foliation $\CZ$. On the other hand, in Section
\ref{sec:7} we have proved that, in the corank 1 case, the control
matrix
$F$ automatically satisfies the condition for \St\ separability,
$N^*dF=F\,dF$. The next proposition shows that, in order to ensure this
condition in the general case, one has to require that the \ham s
$\Ha{a}_i$
be affine \wrt\ the \vefi s $Z_a$.

\begin{prop}
The GZ basis, formed by the $\wHa{a}_i$, is a \stba\ (i.e., it is
\stsep\ in
\dncoo) if and only if $Z_b(Z_c(\Ha{d}_j))=0$ on $S$, for all
$b,c,d=1,\dots,k$ and for all $j=1,\dots,n_d$.
\end{prop}
{\bf Proof.} \St\ separability is equivalent to $N^*dF=F\,dF$, where
$F$ is
the control matrix \rref{eq:g1.10}. Since $dF$
has nonvanishing entries
only in the columns $1,n_1+1,n_2+1,\ldots, n_{k-1}+1$, this condition
takes
the form
\begin{equation}\label{stGZ}
N^*\,d(\wZHa{b}{a}{i})=d(\wZHa{b}{a}{i+1})-\sum_{c=1}^k
\widehat{Z_c(\Ha{b}_i)} d(\wZHa{c}{a}{1})\ ,
\end{equation}
where, as usual, we have put $\Ha{b}_{n_b+1}:=0$. In order to compute
the
left-hand side of \rref{stGZ}, we observe that \rref{secred} implies
\[
P'_S\,df=\left(P'+\sum_{c=1}^k Z_c\wedge X_1^{(c)}\right)dF\ ,
\]
where $f\in C^\infty(S)$ and $F$ is any extension of $f$. Moreover, we have
that
\begin{equation}\label{eq:last}
L_{Z_a}\left(P'+\sum_{c=1}^k Z_c\wedge X_1^{(c)}\right)=
\sum_{c=1}^k [Z_a,Z_c]\wedge X_1^{(c)} =0\ ,
\end{equation}
since the $Z_b$ commute. Hence,
\begin{equation}
\label{stGZb}
\begin{aligned}
P'_S\,&d(\wZHa{b}{a}{i}) =\left(P'+\sum_{c=1}^k Z_c\wedge
X_1^{(c)}\right)d(Z_a(\Ha{b}_i))\\
&= L_{Z_a}\left[\left(P'+\sum_{c=1}^k Z_c\wedge X_1^{(c)}\right)
d\Ha{b}_i\right]= L_{Z_a}\left(P\,d\Ha{b}_{i+1}-\sum_{c=1}^k
Z_c(\Ha{b}_i)P\,
dH_1^{(c)}\right)\\
&= P\,d(Z_a(\Ha{b}_{i+1}))-\sum_{c=1}^k Z_c(\Ha{b}_i)P\,
d(Z_a(H_1^{(c)})) -\sum_{c=1}^k Z_a(Z_c(\Ha{b}_i))P\,
dH_1^{(c)}\ \ ,
\end{aligned}
\end{equation}
so that
\begin{equation}
\label{stGZc}
\begin{aligned}
N^*\,d(\wZHa{b}{a}{i})& = d(\wZHa{b}{a}{i+1})-\sum_{c=1}^k
\wZHa{b}{c}{i}) d(\wZHa{c}{a}{1})\\
&\quad -\sum_{c=1}^k
\widehat{Z_a(Z_c(\Ha{b}_i))}
d\wHa{c}_1\ .
\end{aligned}
\end{equation}
A comparison with \rref{stGZ} completes the proof.
\endpf
Thus, the GZ basis is \stsep\ if (and only if) the second derivatives of the
\ham s along the transversal \vefi s vanish.
This condition is automatically verified in the case of corank $k=1$. This
``discrepancy'' between the generic and the rank $1$ case can be understood
as
follows. Since, by assumption, the transversal distribution $\CZ$ is
integrable, the tubular neighborhood in which it is defined
is equipped with a fibered
structure, in which the fibers are the symplectic leaves of $P$. The
conditions
\[
L_{Z_a}(P)=0;\quad L_{Z_a}\left(P'+
\sum_{c=1}^k Z_c\wedge X_1^{(c)}\right)=0
\]
of equations \rref{zap} and \rref{eq:last} imply that the recursion operator
(to be seen, in this picture, as an endomorphism of the vertical tangent
bundle to the local fibration) is invariant along all the $Z_a$.
So its eigenvalues and hence its minimal polynomial are invariant with
respect to the $Z_a$. In the case $k=1$, as we have seen in Section
\ref{sec:7},
the coefficients of the
minimal polinomial are the derivatives of the Casimir with respect to the
(single) transversal vector field $Z$, but this is not necessarily true in
the
higher corank case.
Notice that, whenever the second derivatives of the Casimirs vanish,
our separated variables are ``invariant''
with respect to the Casimirs, as the
one considered in \cite{Tsig01}.

Still under the assumptions of the above proposition,
the results of
Section \ref{sec:5} tell us how to construct the \St\ matrix and, in
principle, the separation relations. We also know that the entries of
the
\St\ matrix and of the \St\ vector can be used (under additional
hypotheses)
to explicitly find the separation coordinates, i.e., the \dncoo. In the
next
section we will exploit the special properties of the GZ foliation in
order
to determine the
separation relations and, eventually, the \dncoo\ without computing the
\St\
matrix.

\section{Separation relations for GZ systems}
\label{sec:9}

Let us consider the GZ foliation (on the \syml\ $S$) studied in
Subsection
\ref{subsec:8.2}. The aim of this section is to write, in the \stsep\
case,
the separation relations for the \ham s of the GZ basis. To simplify the
notations, we will not use anymore the
symbol $\hat{\ }$ to denote the restriction to $S$.

First of all, we notice that the
relevant information contained in the $n\times n$
control matrix $F$ is actually
encoded in the $k\times k$ polynomial matrix $\Ff(\la)$,
which is the Jacobian matrix
of the Casimirs $\Ha{a}(\la)$ with respect to
the transversal (normalized) vector fields $Z_b$, that is, the
matrix
\begin{equation}
\label{eq:10.1}
\Ff(\la)=\left[ \begin{array}{ccc}
Z_1(\Ha{1}(\la))&\cdots&Z_k(\Ha{1}(\la))\\
\vdots&&\vdots\\
Z_1(\Ha{k}(\la))&\cdots&Z_k(\Ha{k}(\la))\end{array}\right]\>.
\end{equation}
We can translate the results
about separability and \St\ separability
of GZ systems, based on the $n\times n$ matrix equations
\begin{eqnarray}
\label{eq:10.2a} N^* d H &=& F dH\\
\label{eq:10.2b} N^* dF &=& F dF\ ,
\end{eqnarray}
into corresponding equations for
the polynomial matrix $\Ff(\la)$.
To this end we
denote by $\underbar{H}(\la)=(\Ha{1}(\la),\Ha{2}(\la), \ldots,
\Ha{k}(\la))^T $ the $k$-component vector of the polynomial Casimir
functions, and by $\underline{H}_1=(\Ha{1}_1,\Ha{2}_1, \ldots,
\Ha{k}_1)^T$ and $\Ff_1=\left[Z_b(\Ha{a}_1)\right]$ the analogs of the
vector $\underline{H}(\la)$ and of the matrix $\Ff(\la)$, constructed by
using
the coefficients $\Ha{a}_1$ instead of the full Casimir functions
$\Ha{a}(\la)$.
\begin{lem}\label{lem:10.1}
The polynomial control matrix $\Ff(\la)$ satisfies the equation
\begin{equation}
\label{eq:10.3}
(N^*-\la) d\underline{H}(\la)=-\Ff(\la) d \underline{H}_1\>,
\end{equation}
which is the counterpart of the matrix equation~\rref{eq:10.2a}.
\end{lem}
{\bf Proof}.
The $\la^{n_a-i}$-coefficient of the $a$-th row of \rref{eq:10.3} is
exactly
\rref{FforGZ}.
\endpf
In complete analogy, we obtain the ``polynomial form'' of the
St\"ackel separability condition~\rref{eq:10.2b}.
\begin{lem}\label{lem:10.2}
The GZ basis is a St\"ackel basis iff $\Ff(\la)$ satisfies the
condition
\begin{equation}
\label{eq:10.5}
(N^*-\la) d \Ff(\la)=-\Ff(\la) d \Ff_1\>.
\end{equation}
\end{lem}
{\bf Proof}. The simplest way to prove this lemma is to expand both
sides in
powers of $\la$.
We first write \rref{eq:10.5} in componentwise form as
\[
N^* d \Ff_a^b(\la)=\la d \Ff_a^b(\la)-\sum_{c=1}^k \Ff^b_c(\la)d
({\Ff_1})^c_a\>,
\]
and then expand in powers of $\la$, getting
\[
N^* d (Z_a(\Ha{b}_i)=d (Z_a(\Ha{b}_{i+1}))-\sum_{c=1}^k Z_c(\Ha{b}_i)d
(Z_a(\Ha{c}_1))\>,
\]
which are exactly the St\"ackel conditions \rref{stGZ} for the GZ basis.
\endpf

The following lemma shows that the eigenvalues of $N$ can be easily
obtained
from the matrix $\Ff(\la)$.

\begin{lem}\label{lem:10.3}
The determinant of $\Ff(\la)$ is the characteristic
polynomial of $F$. In particular, it coincides with the minimal
polynomial
$\Delta(\la)$ of the recursion operator $N$, that is,
\begin{equation}
\label{eq:11.6}
\det \Ff(\la)=\det(\la I-F)=\Delta(\la)\ .
\end{equation}
\end{lem}
{\bf Proof.} Let $\la_i$ be an eigenvalue of $F$. Then one can check
that
the relative (left) eigenvectors have the form
\[
v_i=(\sig_1^i\la_i^{n_1-1},\sig_1^i\la_i^{n_1-2},\dots,
\sig_1^i,\sig_2^i\la_i^{n_2-1},\dots,
\sig_k^i\la_i^{n_k-1},\dots,\sig_k^i)\ ,
\]
where $\sigma_i:=(\sig_1^i,\dots,\sig_k^i)$ is a nonzero vector such
that
$\sig_i \Ff(\la_i)=0$. This shows that $\det\Ff(\la_i)=0$. Since
$\det\Ff(\la)$ is a monic degree $n$ polynomial and the $\la_i$ are
distinct, we can conclude that \rref{eq:11.6} holds.
\endpf

The next step is to introduce the adjoint (or cofactor) matrix
$\Ff^\vee(\la)$, satisfying the equation
\begin{equation}
\label{eq:11.7}
\Ff^\vee(\la)\Ff(\la)=\Ff(\la)\Ff^\vee(\la)=\det \Ff(\la) I\>.
\end{equation}
We will show that the rows of $\Ff^\vee(\la)$, after a suitable
normalization, provide \stfg s and play the role of the \St\ matrix.
If $\sigma(\la):=e_k\, \Ff^\vee(\la)$ is a row of the adjoint
matrix, then, obviously,
\begin{equation}
\label{eq:11.71}
\sigma(\la)\,\Ff(\la)=\Delta(\la) e_k\>.
\end{equation}
Let $\sigma_j(\la)$ be a nonvanishing entry of $\sigma(\la)$
and let us consider the normalized row
\begin{equation}
\label{eq:11.72}
\rho(\la)=\frac1{\sigma_j(\la)}\sigma\>,
\end{equation}
which satisfies the equation
\begin{equation}
\label{eq:11.73}
\rho(\la)\,\Ff(\la)=\frac{\Delta(\la)}{\sigma_j(\la)} e_k\>.
\end{equation}
\begin{prop}\label{prop:10.1}
Suppose that the component $\rho_a(\la)$ of $\rho(\la)$ is defined for
$\la=\la_i$, $i=1,\dots,n$. Then it is a \stfg, that is, it verifies the
equation
\begin{equation}
\label{eq:10.10}
(N^*-\la)d \rho_a(\la)=0,\quad
\mbox{for $\la=\la_i$, $i=1,\dots,n$}\>.
\end{equation}
\end{prop}
{\bf Proof}.
It is convenient to consider the full vector $\rho(\la)$.
{}From equation~\rref{eq:11.73} we have
\begin{equation}
\label{eq:10.11}
(N^*-\la)d\rho(\la)\cdot \Ff(\la)+ \rho(\la) \cdot (N^*-\la)d \Ff(\la)
=0\quad
\mbox{for $\la=\la_i$} \>.
\end{equation}
Using Lemma~\ref{lem:10.2} we can write the second summand in this
equation
as
\begin{equation}
\label{eq:10.12}
\rho(\la)\cdot (N^*-\la)d \Ff(\la)=- \rho(\la)\Ff(\la) d \Ff_1\quad
\mbox{for
$\la=\la_i$,}
\end{equation}
so that we finally obtain
\begin{equation}
\label{eq:10.13}
(N^*-\la_i)d\rho(\la_i)\cdot \Ff(\la_i)=0\quad
\mbox{for $i=1,\dots,n$.}
\end{equation}
But the kernel of $\Ff(\la_i)$ is 1-dimensional, due to the fact that
the
$\la_i$ are distinct.

Indeed, from \rref{eq:11.7} we have that
\begin{equation}
\label{detagg}
\det\Ff^\vee(\la)=(\det \Ff(\la))^{k-1}=\prod_{i=1}^n (\la-\la_i)^{k-1}\
.
\end{equation}
If $\dim\mbox{ker}\,\Ff(\la_i)\ge 2$ for some $i$, then the rank of
$\Ff(\la_i)$ would be less than $k-1$, so that $\Ff^\vee(\la_i)=0$,
and therefore
$\Ff^\vee(\la)=(\la-\la_i)\tilde\Ff(\la)$ for some polynomial
matrix $\tilde\Ff(\la)$. But then $\det\Ff^\vee(\la)=
(\la-\la_i)^{k}\det \tilde\Ff(\la)$, contradicting \rref{detagg}.

Coming back to \rref{eq:10.13}, we can assert that
there exist 1-forms $\nu_i$ such that
\begin{equation}
\label{eq:10.14}
(N^*-\la_i)d\rho(\la_i)=\nu_i\rho(\la_i)\quad
\mbox{for $i=1,\dots,n$.}
\end{equation}
Since the $j$-th component of $\rho(\la_i)$ is 1, we have that
the $\nu_i$ vanish, and this closes the proof.
\endpf
Now we are ready to show how to compute the separation equations for GZ
systems.
\begin{prop}\label{prop:10.2}
Let the $\rho_a(\la)$ be as in the previous proposition and suppose that
they are defined for $\la=\la_i$. Then
$\sum_{a=1}^k \sigma_a(\la)\Ha{a}(\la)$ is a \stfg.
\end{prop}
{\bf Proof}. Let us write compactly
$\sum_{a=1}^k \rho(\la)\Ha{a}(\la)=
\rho(\la)\cdot \underline{H}(\la)$ and compute
\begin{equation}
\label{eq:10.15}
(N^*-\la)d (\rho(\la)\cdot \underline{H}(\la)) =
(N^*-\la)d\rho(\la)\cdot \underline{H}(\la)+
\rho(\la)(N^*-\la)d\underline{H}(\la)\>.
\end{equation}
For $\la=\la_i$ the first summand vanishes thanks to
Proposition~\ref{prop:10.1}, while the
second equals (according to Lemma~\ref{lem:10.1})
\[
-\rho(\la_i)\, \Ff(\la_i)\, d\underline{H}_1\ ,
\]
and so vanishes as well.
\endpf
Therefore we have shown that the separation relations of the
GZ basis (in the St\"ackel case) are given by
\begin{equation}
\label{eq:10.17}
\sum_{a=1}^k \rho_a(\la_i)\Ha{a}(\la_i)=\Phi_i(\la_i,\mu_i)\ ,
\quad i=1,\ldots, n\ ,
\end{equation}
that in the corank 1 case boils down to
equation \rref{seprelco1}.

We end this section with the following remark. Let us suppose that the
multipliers $\rho_a(\la)$ and the coordinates $\mu_1,\dots,\mu_n$ be
related by a ``simple'' algebraic expression, e.g., that there exist
integer numbers $p_1=0$, $p_2$,\dots,$p_k$ such that
\[
\mu_i^{p_a}=\rho_a(\la_i)\ ,\qquad i=1,\dots,n\mbox{ and }
a=1,\dots,k\ .
\]
This means, according to the results of Section \ref{sec:6}, that the
$p_a$--th root $\rho_a(\la)$ is a \stfg\ satisfying the equation
\rref{eq:nmu}, i.e.,  $Y(\sqrt[p_a]{\rho_a(\la)})=1$, for $\la=\la_i$.
Then the separation relations \rref{eq:10.17} ``degenerate'' to a
single one, that is, they can be read as the vanishing of the
two-variable function
\begin{equation}
\label{eq:10.17b}
\sum_{a=1}^k \mu^{p_a} \Ha{a}(\la)-\Phi(\la,\mu)\
\end{equation}
evaluated at the points $(\la_i,\mu_i)$, for $i=1,\dots,n$. Hence, in
such an instance, we can associated with the GZ system a ``spectral
curve'' over which the separation coordinates lie.

This is an indication, which is verified in several concrete examples,
that the theory herewith presented may provide an effective bridge
between the classical theory of the Hamilton-Jacobi equation and its
modern outsprings, related to algebraic integrability. In this
respect, several questions naturally arise, namely,
\begin{enumerate}
\item Can the degeneration property of the separation relations be
characterized in terms of the \bih\ structure?
\item In this case, what can one say about the algebraicity of the
separation relation?
\end{enumerate}
We will further address these problems in \cite{fmp2}.
In this paper we limit
ourselves to give an example related to loop algebras, where all these
features are present.

\section{An example related to $\mathfrak{sl}(3)$}
\label{sec:10}

Applications of the scheme we have presented in this paper have already
appeared in the literature. Namely, in~\cite{FMT00} a preliminary
picture of
these ideas has been applied to the $t_5$--stationary reduction of
the Boussinesq \ger y. Subsequently, in \cite{FMPZ2} we have shown how
to
frame
all stationary reductions of the KdV theory inside this picture, and in
\cite{creta} the classical $A_n$--Toda lattices have been considered
(see also \cite{cambri} for the Neumann system, and \cite{DM}).
In this final section we will illustrate how our theoretical scheme
concretely works in an example, which is related to the
$t_5$--stationary Boussinesq system, in the sense that the latter can
be obtained via a reduction from the one we will present.
Even if,
for the sake of brevity, we will stick to such a particular example, we
claim that the same arguments hold for a wide class of integrable
systems on
finite-dimensional orbits of loop algebras, studied, e.g., in
\cite{Ha1}. In these
cases, the DN (separation) coordinates turn out to be
the so-called spectral Darboux coordinates \cite{Ha1,Ha2}.

The system we are going to study is defined on the space
$\mathfrak{sl}(3)\times\mathfrak{sl}(3)$ of pairs $(X_0,X_1)$ of
$3\times 3$
traceless matrices. The cotangent (and the tangent) space at a point is
identified with the manifold itself via the pairing
\[
\langle (V_0,V_1),(W_0,W_1)\rangle =\tr(V_0W_0+V_1W_1)\ ,
\]
so that the differential of a scalar function $F$
is represented by a pair of matrices,
\[
dF=\left(\ddd{F}{X_0},\ddd{F}{X_1}\right)\ .
\]
We introduce \cite{MM86,RSTS88} the two compatible Poisson tensors
defined,
at the point $(X_0,X_1)$, by
\[
\begin{array}{c}
P:\quad
\left[\begin{array}{c} {V_0}\\ {V_1}\end{array}\right]
\mapsto
\left[\begin{array}{c} [X_1,V_0]+[A,V_1] \\
{[A,V_0]}
\end{array}\right]\\
\noalign{\vspace{1truemm}}\\
P':\quad
\left[\begin{array}{c} {V_0}\\ {V_1}\end{array}\right]
\mapsto
\left[\begin{array}{c} -[X_0,V_0]\\ {[A,V_1]}
\end{array}\right]\ ,
\end{array}
\]
where
\[
A=\left[ \begin{array}{ccc}
0&0&0\\
1&0&0\\
0&1&0
\end{array}\right]\ .
\]
One can easily see that the functions
\begin{eqnarray*}
C_1(X_0,X_1)&=&\tr(AX_1)=(X_1)_{12}+(X_1)_{23}\\
C_2(X_0,X_1)&=&\tr(A^2X_1)=(X_1)_{13}
\end{eqnarray*}
are common Casimirs of $P$ and $P'$. Thus the \bih\ structure can be
trivially restricted to
\[
M=\{(X_0,X_1)\in \mathfrak{sl}(3)\times\mathfrak{sl}(3)\mid
(X_1)_{12}+(X_1)_{23}=0\ ,\ (X_1)_{13}=1\}\ ,
\]
which is the 14-dimensional manifold where our GZ system will be
defined.
Indeed, it can be directly shown (see also \cite{RSTS88}) that, if
\[
L(\la)=\la^2 A+\la X_1+X_0\ ,
\]
then
\begin{equation}
\label{eq:ex7}
\Ha{1}=\frac{1}{2}\tr L(\la)^2\quad\mbox{and}\quad
\Ha{2}=\frac{1}{3}\tr L(\la)^3
\end{equation}
are Casimir functions of the Poisson pencil $P_\la=P'-\la P$. One finds
that
\begin{equation}
\label{casex}
\begin{split}
&\Ha{1}=\la^3+\Ha{1}_0\la^2+\Ha{1}_1\la+\Ha{1}_2\\
&\Ha{2}=\la^5+\Ha{2}_0\la^4+\Ha{2}_1\la^3+\Ha{2}_2\la^2
+\Ha{2}_3\la+\Ha{2}_4\ ,
\end{split}
\end{equation}
where
\[
\begin{array}{c}
\Ha{1}_0=\tr(AX_0+\frac12 {X_1}^2)\quad
\Ha{1}_1=\tr(X_0X_1)\quad
\Ha{1}_2=\frac12 \tr {X_0}^2 \\
\Ha{2}_0=\tr(A^2X_0+A{X_1}^2)\quad
\Ha{2}_1=\tr(\frac13 {X_1}^3+AX_0X_1+AX_1X_0) \\
\Ha{2}_2=\tr({X_1}^2 X_0+A {X_0}^2)\quad
\Ha{2}_3=tr(X_1{X_0}^2)\quad
\Ha{2}_4=\frac13\tr{X_0}^3\ .
\end{array}
\]
Obviously, $\Ha{1}_0$ and $\Ha{2}_0$ are Casimirs of $P$, whereas
$\Ha{1}_2$ and $\Ha{2}_4$ are Casimirs of $P'$. Since the differentials
of
the functions $\Ha{a}_i$ are linearly independent on a dense open
subset of $M$, and the corank of
$P$ and $P'$ is 2,
we can conclude that the hypotheses of Section \ref{sec:8} are
verified, with $k=2$, $n_1=2$, and $n_2=4$.
The GZ system on $M$ is given by the 6 \bih\ \vefi s associated with the
coefficients of the Casimirs \rref{casex}. The first \vefi s of the two
\bih\ hierarchy are, respectively,
\begin{equation}
\label{fvfex}
X_1^{(1)}=([A,X_0],[A,X_1])\ ,\qquad
X_1^{(2)}=([A^2,X_0],[A^2,X_1])\ .
\end{equation}

Let us fix a \syml\ $S$ of $P$, defined by the constraints
$\Ha{1}_0=c^1$, $\Ha{2}_0=c^2$. According to Proposition
\ref{prop:AL-inte}, the 6 remaining \ham s define a Lagrangian
foliation,
called the GZ foliation, on the 12-dimensional \syman\
$S$. The results of Section \ref{sec:8} entail that, in order to
separate
the GZ system, we need a distribution which is transversal to the
symplectic
leaves of $P$. More precisely, let $\parpu$ and $\parpu'$ be the \parp s
associated with $P$ and $P'$. Then we must find a pair of vector fields
$(Z_1,Z_2)$, spanning a 2-dimensional integrable distribution on $M$,
such
that
\begin{equation}
\label{normex}
Z_a(\Ha{b}_0)=\delta_a^b
\end{equation}
and such that the functions invariant along the distribution form
a Poisson subalgebra \wrt\ both $\parpu$ and $\parpu'$. It is not
difficult
to show that these requirements are fulfilled by
\begin{eqnarray*}
& Z_1 &:\qquad {\dot X}_0=E_{23}\ ,\quad {\dot X}_1=0\\
& Z_2 &:\qquad {\dot X}_0=E_{13}\ ,\quad {\dot X}_1=0\ ,
\end{eqnarray*}
where $E_{ij}$ is the matrix with 1 in the $(i,j)$ entry and 0
elsewhere. In
fact, a function $F\in C^\infty(M)$ is invariant \wrt\ both $Z_1$ and
$Z_2$
if and only if
\begin{equation}
\label{parab}
\left(\ddd{F}{X_0}\right)_{31}=\left(\ddd{F}{X_0}\right)_{32}=0\ ,
\end{equation}
and such functions form a Poisson subalgebra, because the matrices
fulfilling \rref{parab} are a Lie subalgebra of $\mathfrak{sl}(3)$.
Moreover,
$Z_1(\Ha{1}_0)=\tr(AE_{23})=1$ and
$Z_1(\Ha{2}_0)=\tr(A^2 E_{23})=0$, and the analogous equations
for $Z_2$ hold, so that we have the normalization \rref{normex}. Then
Proposition \ref{prop:zvefi} implies that
\[
L_{Z_a}P=0\ ,\qquad L_{Z_a}P'=\sum_{b=1}^2 Y_a^b\wedge Z_b\ ,
\]
with $Y_a^b=[Z_a,X_1^{(b)}]$. The \vefi\
(see Remark \ref{rem:yvf})
\[
Y=Y_1^1+Y_2^2=[Z_1,X_1^{(1)}]+[Z_2,X_1^{(2)}]
\]
is given, on account of \rref{fvfex}, by
\[
{\dot X}_0=
\left[ \begin{array}{ccc}
-1&0&0\\
0&-1&0\\
0&0&2
\end{array}\right]
\ ,\quad {\dot X}_1=0\ .
\]
Hence, the \syml\ $S$ has an $\om N$ structure and Theorem
\ref{teo:gzsep} tells us that the above-defined GZ foliation is
separable in \dncoo.

Now we will use the
results of Section \ref{sec:8} and \ref{sec:9} to discuss the
St\"ackel separability and the separation relations of the GZ basis.
Indeed, $\Ha{1}(\la)$ and $\Ha{2}(\la)$ are easily seen to be affine
\wrt\
the transversal \vefi s,
\[
Z_a(Z_b(\Ha{1}(\la)))=Z_a(Z_b(\Ha{2}(\la)))=0\qquad \mbox{for all
$a,b=1,2$,}
\]
meaning that the GZ basis is St\"ackel separable.

A set of special \dncoo\
$(\la_i,\mu_i)_{i=1,\dots,6}$ on $S$ is determined as follows.
We write the polynomial matrix
\[
\Ff(\la)=\left[
\begin{array}{cc}
Z_1(\Ha{1}(\la)) & Z_2(\Ha{1}(\la)) \\
Z_1(\Ha{2}(\la)) & Z_2(\Ha{2}(\la))
\end{array}
\right]
=\left[
\begin{array}{cc}
L(\la)_{32} & L(\la)_{31} \\
(L(\la)^2)_{32} & (L(\la)^2)_{31}
\end{array}
\right]\ ,
\]
whose determinant gives the minimal polynomial of the recursion operator
$N$
of $S$:
\begin{equation}
\label{exlam}
\det \Ff(\la)=\la^6-\sum_{i=1}^6 c_i\la^{6-i}\ .
\end{equation}
Its roots are the eigenvalues $(\la_1,\dots,\la_6)$ of $N$. The $\mu_i$
coordinates can be found with the strategy described in Section
\ref{sec:6},
which consists in looking for a \stfg\ $f(\la)$ such that $Y(f(\la))=1$.
In
Section \ref{sec:9} we saw that the normalized rows of the adjoint
matrix
$\Ff^\vee(\la)$ of $\Ff(\la)$ provide \stfg s. We have
\[
\Ff^\vee(\la)=\left[
\begin{array}{cc}
(L(\la)^2)_{31} & -L(\la)_{31} \\
-(L(\la)^2)_{32} & L(\la)_{32}
\end{array}
\right]\ ,
\]
so that $f(\la):=-(L(\la)^2)_{31} / L(\la)_{31}$ is a \stfg. Since
\[
Y((L(\la)^2)_{31})=-L(\la)_{31}\mbox{ and } Y(L(\la)_{31})=0\ ,
\]
we obtain
$Y(f(\la))=1$, and therefore
\begin{equation}
\label{exmu}
\mu_i=f(\la_i)=-(L(\la_i)^2)_{31} / L(\la_i)_{31}\ ,\quad
i=1,\dots,6\ ,
\end{equation}
form with the $\la_i$ a set of special \dncoo.

At this point we could, in principle,
use \rref{exlam} and \rref{exmu} to explicitly write
the point $(X_0,X_1)$ of $S$ in terms of $(\la_i,\mu_i)_{i=1,\dots,6}$,
and
we could compute the functions $\Phi_i$ in
\rref{eq:10.17} in order to obtain the separation relations for the GZ
basis:
\[
\rho_1(\la_i)\Ha{1}(\la_i)+\rho_2(\la_i)\Ha{2}(\la_i)
=\Phi_i(\la_i,\mu_i)\ ,
\]
with $\rho_1(\la)=f(\la)$ and $\rho_2(\la)=1$. Thus we have
\begin{equation}
\label{exseprel}
\mu_i \Ha{1}(\la_i)+\Ha{2}(\la_i)=\Phi_i(\la_i,\mu_i)\ .
\end{equation}
However, we can directly show that these separation relations
coincide with the ones
given by the spectral curves, i.e.,
\[
\det(\mu I-L(\la))=0\ .
\]
Since $\det(\mu I-L(\la))=\mu^3-\frac12\tr(L(\la)^2)\mu
-\frac13\tr(L(\la)^3)
%=\mu_3-\mu\Ha{1}-\Ha{2}
$, the points
$(\la_i,\mu_i)_{i=1,\dots,6}$ given by \rref{exlam} and \rref{exmu}
belong
to the spectral curve if and only if
\begin{equation}
\label{exspc}
-\left(\frac{(L(\la_i)^2)_{31}}{L(\la_i)_{31}}\right)^3+
\frac12\tr(L(\la_i)^2)\frac{(L(\la_i)^2)_{31}}{L(\la_i)_{31}}
-\frac13\tr(L(\la_i)^3)=0
\end{equation}
for all $\la_i$ such that
\begin{equation}
\label{exlambis}
L(\la_i)_{32}(L(\la_i)^2)_{31} - L(\la_i)_{31}
(L(\la_i)^2)_{32}=0\ .
\end{equation}
Since it can be checked that equation \rref{exspc} holds for every
traceless
$3\times 3$ matrix fulfilling \rref{exlambis}, we have
indeed shown that the
separation relations \rref{exseprel} are given by the spectral curve.

\end{document}